\documentclass[floatfix,aps,prd,eqsecnum,showpacs,amsmath,amssymb,nofootinbib,preprintnumbers]{revtex4}

\usepackage{graphicx}
\usepackage{color}

\usepackage{hyperref}
\hypersetup{colorlinks=true,linkcolor=blue,pdfstartview=XYZ}

\def\laq{~\raise 0.4ex\hbox{$<$}\kern -0.8em\lower 0.62ex\hbox{$\sim$}~}
\def\gaq{~\raise 0.4ex\hbox{$>$}\kern -0.7em\lower 0.62ex\hbox{$\sim$}~}

\def\beq{\begin{equation}}
\def\eeq{\end{equation}}
\def\bea{\begin{eqnarray}}
\def\eea{\end{eqnarray}}

\def \pa {\partial}
\def \ra {\rightarrow}
\def \ti {\widetilde}

\def \ga {\gamma}

    \def\be{\begin{equation}}
    \def\ee{\end{equation}}
    \def\ba{\begin{eqnarray}}
    \def\ea{\end{eqnarray}}

\newcommand{\eq}{\begin{equation}}
\newcommand{\eqx}{\end{equation}}
\newcommand{\eqn}{\begin{eqnarray}}
\newcommand{\eqnx}{\end{eqnarray}}

\newcommand{\Ups}{\Upsilon}

\newcommand{\rref}[1]{(\ref{#1})}

\newcommand{\Acal}{\mathcal A}

\newcommand{\Hcal}{\mathcal H}

\usepackage{pifont}

\begin{document}

\preprint{CERN-PH-TH/2012-225}
\preprint{LPTENS-12/27}
\preprint{NSF-KITP-12-178}

\title{The second-order  luminosity-redshift relation in a generic inhomogeneous cosmology}

\author{Ido Ben-Dayan$^{1,2}$,  Giovanni Marozzi$^{3,4}$, Fabien Nugier$^{5}$ and Gabriele Veneziano$^{3,6}$}

\affiliation{$^1$Canadian Institute for Theoretical Astrophysics, 60 St
George, Toronto ON, M5S 3H8\\
$^{2}$ Perimeter Institute for Theoretical Physics,
Waterloo, Ontario N2L 2Y5, Canada\\
 $^{3}$ Coll\`ege de France, 11 Place M. Berthelot, 75005 Paris, 
 France\\
 $^{4}$ Universit\'e de Gen\`eve, D\'epartement de Physique Th\'eorique and CAP,
24 quai Ernest-Ansermet, CH-1211 Gen\`eve 4, Switzerland \\
$^{5}$ Laboratoire de Physique Th\'eorique de l'\'Ecole Normale Sup\'erieure, CNRS UMR 8549, 24 Rue Lhomond, 75005 Paris, France\\
$^{6}$CERN, Theory Unit, Physics Department, CH-1211 Geneva 23, Switzerland}



\begin{abstract}
After recalling a general non-perturbative expression for the luminosity-redshift relation holding in a recently proposed ``geodesic light-cone" gauge, we show how it  can be transformed to phenomenologically more convenient gauges in which cosmological perturbation theory is better understood.
We present, in particular,  the complete result on the luminosity-redshift relation in the Poisson gauge up to second order for a fairly generic perturbed cosmology, assuming that appreciable vector and tensor perturbations are only generated at second order. 
This relation provides a basic ingredient for the computation of the effects of stochastic inhomogeneities on precision dark-energy cosmology  whose results we have anticipated in a recent letter. More generally, it can be used in connection with  any physical information carried by light-like signals traveling along our past light-cone.
\end{abstract}

\vspace {1cm}~


\maketitle
\section {Introduction and Outline}
\label{Sec1}
\setcounter{equation}{0}

In a recent letter \cite{BGMNV2} (see also \cite{BGMNV1}) we have computed the effects of a stochastic background of inhomogeneities on the determination of dark-energy parameters in precision cosmology. The outcome of that analysis has been that such perturbations cannot simulate a substantial fraction of dark energy: indeed, their contribution to the averaged flux-redshift relation is both too small (especially at large redshift) and has the wrong $z$-dependence.  Nonetheless, stochastic fluctuations add a new and relatively important dispersion with respect to the prediction of the homogeneous and isotropic Friedmann-Lema\^itre-Robertson-Walker (FLRW) cosmology. This dispersion is independent of the experimental apparatus, the observational procedure, or the dispersion in absolute luminosity.  Given the present (and probably near-future) limited statistics of Supernovae data, this, together with other phenomena, may prevent a determination of $\Omega_{\Lambda}(z)$ down to the percent level using the luminosity-redshift relation alone.

In  \cite{BGMNV2} we have presented the main ideas  and  most significant results of the calculation which, essentially, proceeds in two successive steps. The first one is the computation of the luminosity  redshift relation $d_L(z)$ (or, equivalently, of the flux $\sim d_L^{-2}$) at second order in perturbation theory.  
The method used in \cite{BGMNV2}, being gauge invariant \cite{GMNV}, allows us to express the result in a convenient gauge in which perturbations are known to second order, the so-called Poisson gauge (PG) \cite{PG}~\footnote{Following the pioneering work of \cite{Sasaki}, $d_L$ has been already computed to first order in the longitudinal gauge (for a CDM model in \cite{Bonvin}, CDM and $\Lambda$CDM in \cite{Pyne:2003bn}
and for a generic model in \cite{BGMNV1}), and to second order in the synchronous gauge, but only for a dust-dominated Universe, in \cite{Barausse:2005nf}.}.
The second step consists of performing the relevant light-cone/ensemble averages, as in \cite{BGMNV1}, and in inserting a realistic power spectrum of stochastic perturbations. This gives their effect on dark-energy parameters at the quantitative level.

In this paper we will present full details about the first stage of this two-step process leaving the details about the second one to a future publication \cite{BGMNV4}. One reason for doing so is that the calculation of $d_L$ is independent of the rest of the calculation, has an interest of its own (i.e. irrespectively of its subsequent application to light-cone/ensemble averaging) and could possibly find many other applications in precision cosmology. Furthermore, the result presented here for $d_L$ is valid in general, i.e. for any given background model~\footnote{Except if caustics form. It has been argued \cite{Ellis:1998} that the area distance is modified when caustics are present inside the past light-cone.}.  

The paper is organized as follows.
In Section 2 we specify the PG 
up to second order in perturbation theory. We then recall
 the definition and special properties of 
an adapted system of coordinates introduced in \cite{GMNV}
dubbed the geodesic light-cone (GLC) gauge.
 We also give the connection between the two gauges up to second order.
In Section 3 we present the actual calculation of $d_L$ and express the final result, in compact form, in terms of  perturbations in the PG, of the observed redshift, and of the observer's angular coordinates. The long, full expression of  $d_L$ can be found in the Appendix together with a recollection of  our definitions. In Section 4 we first offer some physical interpretation of the various terms appearing in $d_L$ and then show how the final result can be averaged over the observer's past light-cone reproducing the formulae used in  \cite{BGMNV2}.
In Section 5 we summarize the results and draw some short conclusions.

We note that, after we submitted our short paper \cite{BGMNV2} -- and while preparing this one -- another group \cite{Umeh:2012pn} has submitted a summary of their own calculation of $d_L$ in the PG and for a $\Lambda$CDM model. Since their calculation and ours are very different (and obviously completely independent) comparing the final outcomes, for such particular case of a $\Lambda$CDM model, will provide a very useful test of this highly non-trivial, long and somewhat tricky calculation.


\section{From the Poisson  to the Geodesic Light-Cone gauge at second order}
\label{Sec2}
\setcounter{equation}{0}


\subsection{The Poisson gauge}

Let us consider a non-homogeneous space-time approximated by a spatially flat FLRW Universe plus scalar, vector and tensor perturbations. 
In the so-called Poisson gauge (PG) (\cite{PG}), a generalization of the Newtonian (or longitudinal) gauge beyond first order,  the corresponding metric takes the following  standard form in cartesian coordinates:
\be
ds_{PG}^2 = a^2(\eta) \left( -(1+ 2 \Phi) d\eta^2 + 2 \omega_i d\eta d x^i + \left[(1- 2 \Psi)\delta_{ij} + h_{ij}   \right] dx^i dx^j \right) ~~,
\label{PGmetricstandard}
\ee
where $\Phi$ and $\Psi$ are scalar perturbations, 
$\omega_i$ is a transverse vector ($\partial^i \omega_i=0$) and $h_{ij}$ is a transverse and traceless
tensor ($\partial^i h_{ij}=0=h^i_i$).
This metric depends on six arbitrary functions, hence it is completely gauge fixed. Up to second order  the (generalized) Bardeen potentials $\Phi$ and $\Psi$ are defined as follows:
\be
\Phi \equiv \psi + \frac{1}{2} \phi^{(2)} ~~,~~ \Psi \equiv \psi + \frac{1}{2} \psi^{(2)} ~~,
\ee
where we have assumed no anisotropic stress in order to set $\Psi = \Phi= \psi$ at first order. In this paper we shall consider $\omega_i$ and $h_{ij} $ as second order quantities, the idea being that, in inflationary cosmology, first order scalar perturbations dominate over the others for small slow-roll parameters. On the other hand vector and tensor perturbations are automatically generated from scalar perturbations at second order (see e.g. \cite{Matarrese:1997ay,BMR}).


\subsection{The geodesic light-cone gauge}

For problems associated with the observation of light sources lying on the past light-cone of a given observer, it is convenient to identify the null hypersurfaces on which the photons reach the observer with those on which a null coordinate takes constant values. For this reason  we have  introduced in \cite{GMNV}  an adapted system of coordinates -- defining what  we have called a  ``geodesic light-cone'' (GLC) gauge -- in which several quantities greatly simplify \cite{GMNV}  while keeping all the required degrees of freedom for applications to general geometries. 

Let's  recall \cite{GMNV} that the coordinates $x^\mu= (\tau, w, \ti{\theta}^a)$  (with $a=1,2$, $\ti{\theta}^1 = \ti{\theta}$, $\ti{\theta}^2 = \ti{\phi}$) specifying the metric in the GLC gauge correspond to a complete gauge fixing of the so-called observational coordinates, defined in \cite{Maartens1,Maartens3,Clarkson}. The GLC metric too depends on six arbitrary functions 
($\Ups$, a two-dimensional ``vector" $U^a$ and a symmetric matrix $\gamma_{ab}$),
and its line element takes the form 
\bea
\label{LCmetric}
ds_{GLC}^2 = \Ups^2 dw^2-2\Ups dw d\tau+\gamma_{ab}(d\ti{\theta}^a-U^a dw)(d\ti\theta^b-U^b dw) ~~.
 \eea
In matrix form,  the metric and its inverse read:
\eq
\label{GLCmetric}
g^{GLC}_{\mu\nu} =
\left(
\begin{array}{ccc}
0 & - \Ups &  \vec{0} \\
-\Ups & \Ups^2 + U^2 & -U_b \\
\vec0^{\,T}  &-U_a^T  & \gamma_{ab} \\
\end{array}
\right) ~~ ~~~~~,~~ ~~~~~
g_{GLC}^{\mu\nu} =
\left(
\begin{array}{ccc}
-1 & -\Ups^{-1} & -U^b/\Ups \\
-\Ups^{-1} & 0 & \vec{0} \\
-(U^a)^T/\Ups & \vec{0}^{\, T} & \gamma^{ab}
\end{array}
\right) ~~,
\eqx
where $\vec 0=(0,0)$, $U_b= (U_1, U_2)$, while the $2 \times 2$ matrices 
$\gamma_{ab}$ and  $\gamma^{ab} = (\gamma_{ab})^{-1} $ lower and  raise  the two-dimensional indices. Clearly $w$ is a null coordinate (i.e.  $\pa_\mu w \thinspace \pa^\mu w=0$), and a past light-cone hypersurface is specified by the condition $w=$ constant. We can also easily check that $\partial_{\mu} \tau$ defines a geodesic flow, i.e. that
$\left( \pa^\nu \tau\right) \nabla_\nu \left( \pa_\mu \tau\right) = 0$
(as a consequence of the relation $g^{\tau \tau} = -1$). 

In the limiting case  of  a spatially flat homogeneous FLRW geometry, with scale factor $a$, cosmic time $t$, and conformal time parameter $\eta$ such that $d\eta= dt/a$, the transformations to the GLC coordinates and the  meaning of the new metric components are easily found as follows \cite{GMNV}:
\bea
&&
\tau=t ~,  ~~~~~~~~~~~~~~~ w= r+\eta ~,~~~~~~~~~~~
\Ups = a(t) ~,
\nonumber \\ &&
	U^a=0 ~, ~~~~~~  \gamma_{ab}d\ti{\theta}^a d\ti\theta^b = a^2(t) r^2 (d\ti{\theta}^2 +\sin^2 {\ti\theta} d\ti{\phi}^2) \equiv \gamma_{ab}^{FLRW} d\ti{\theta}^a d\ti\theta^b ~.
\label{FR}
\eea

Even though we will be mainly using the GLC gauge for a perturbed FLRW metric in the PG,  it is important to stress that it is always possible to choose the GLC coordinates in such a way that $\tau$ and $t$ 
of the synchronous gauge are identified like in the above homogeneous FLRW limit  \cite{BGMNV1}.
As a consequence we can easily introduce with $\tau$ a family of geodetic reference observers which exactly coincide with the static ones of the synchronous gauge. We also remark that, in GLC coordinates, the null geodesics connecting sources and observer are characterized by the simple tangent vector $k^{\mu} = g^{\mu \nu} \partial_{\nu} w =  g^{\mu w} = - \delta^{\mu}_{\tau} \Ups^{-1}$, meaning that photons  travel at constant $w$ and $\ti \theta^a$. This makes the calculation of  the redshift and of the area distance particularly easy in this gauge.

Let us denote by the subscripts ``o'' and ``s'', respectively, a quantity evaluated at the observer and source space-time position, and consider a light ray emitted by a static geodetic source lying at the intersection between the past light-cone of a static geodetic observer (defined by the equation $w=w_o$) and the spatial hypersurface $\tau= \tau_s$ with $\tau_s$ taken momentarily as a constant. The light ray will be received by such static geodetic observer at $\tau=\tau_o>\tau_s$. The redshift $z_s$ associated with this  light ray is then given by \cite{GMNV}:
\be
\label{redshift}
(1+z_s) = \frac{(k^{\mu} u_{\mu})_s }{(k^{\mu} u_{\mu})_o}  = \frac{(\partial^{\mu}w \pa_\mu \tau)_s }{(\partial^{\mu}w \pa_\mu \tau)_o}  = {\Ups(w_o, \tau_o, \ti \theta^a)\over \Ups(w_o, \tau_s, \ti \theta^a)} ~~.
\ee
We will denote by $\Sigma(w_o, z_s)$ the two-dimensional surface (topologically a sphere) which lies on our past light-cone ($w=w_o$)  and corresponds to a fixed  redshift ($z = z_s$). In terms of the $\tau$ coordinate this will correspond to  imposing  the equation $\tau = \tau_s(\tilde{\theta}^a, w_o, z_s)$ enforcing (\ref{redshift}). Hereafter $\tau_s$ will denote this (in general angle-dependent) quantity.

As said, also the area distance $d_A$, related to the luminosity distance $d_L$ of a source 
at redshift $z_s$ by the Etherington  (or reciprocity) relation \cite{Et1933}
\be
 d_A = (1+z_s)^{-2} d_L ~~,
 \label{dLdA}
\ee
 takes a particularly simple form in the GLC gauge \cite{GMNV}. We begin recalling the definition of $d_A$ \cite{KS}
\be
\label{KS}
d_A^2 \equiv \frac{dS}{d\Omega_o} ~~,
\ee
where $d\Omega_o$ is the infinitesimal solid angle at the observer, and $dS$ is the cross-sectional area element perpendicular to the light ray at the source. Let us then show that, in the GLC gauge, we have  \cite{GMNV}:
\be
d_A^2 = \frac{\sqrt{\gamma}}{\sin \tilde{\theta}} ~~.
\label{Eq_dA}
\ee
Indeed, $\gamma_{ab}$ is nothing but the induced metric on the surface $\Sigma(w_o, z_s)$  provided this is parametrized in terms of the two ``world-sheet" coordinates $\xi^a \equiv  \tilde{\theta}^a $ and otherwise  given by $w = w_o$, $\tau = \tau_s(\tilde{\theta}^a, w_o, z_s)$. Using the standard definition of an induced metric:
\be
\label{gammaab}
\gamma_{a b}^{ind} = \frac{\partial x^{\mu}}{\partial \xi^{a}} \frac{\partial x^{\nu}}{\partial \xi^{b}} g_{\mu \nu}(x) ~~,
\ee
manifestly independent of the spacetime coordinates $x^{\mu}$ being used, we simply find:
\be
\label{gamma=gammaab}
\gamma_{ab}^{ind} = \ga_{ab} ~~,
\ee
since $w = w_o$ (independently of $\tilde{\theta}^a $) and
the only non-zero entry for the metric with a lower index $\tau$ is $g_{\tau w}$.
We can also argue that the area element computed on this surface is orthogonal to the null geodesics and can therefore be identified with the $dS$ of (\ref{KS}).
Indeed, consider the projection of the photon momentum along the constant-$z_s$ hypersurface  (which in this gauge, by (\ref{redshift}), corresponds to constant $\Ups$): 
\be
k_{\mu \|} = k_{\mu} - \frac{k_{\nu} \partial^{\nu} \Ups}{\partial_\sigma \Ups \partial^\sigma \Ups} \partial_{\mu} \Ups ~~,
\ee 
and the particular linear combination 
\be
n_{\mu}^{(z_s)} =  \alpha ~ \partial_{\mu} w + \beta ~ \partial_{\mu} \Ups \, ,
\ee
defining the normal to $\Sigma$ lying on the same constant-$z_s$ hypersurface and thus satisfying:
\be
n_{\mu}^{(z_s)}\partial^{\mu} \Ups = 0 \Rightarrow (\partial_{\tau} \Ups)~ \alpha = \Ups (\partial_{\mu} \Ups \partial^\mu \Ups)~ \beta \, .
\ee
One  can easily verify that $n_{\mu}^{(z_s)}$ is exactly parallel to $k_{\mu \|}$. 
Finally, using their constancy along the null geodesics, we can also identify $\tilde{\theta}^a$ with  the angular coordinates at the observer's position where, within an infinitesimal region, we can take the metric to be flat. Therefore, as promised,
\be
\label{dAGLC}
d_A^2 = \frac{dS}{d\Omega_0} = \frac{d^2 \tilde{\theta} \sqrt{\gamma}}{d^2 \tilde{\theta} \sin \tilde{\theta}} = \frac{\sqrt{\gamma}}{\sin \tilde{\theta}} ~~.
\ee

The above expressions for the area distance $d_A$ singles out the flux $\Phi \sim d_L^{-2} = (1+z_s)^{-4} d_A^{-2}$ as an  important, and extremely simple, observable to average over the 2-sphere $\Sigma(w_o, z_s)$ embedded in the light-cone:
\bea 
\label{Phitheory}
\langle d_L^{-2} \rangle(w_o,z_s)&=&(1+z_s)^{-4}\frac{\int dS \frac{d\Omega_0}{dS}}{\int dS}=(1+z_s)^{-4}\frac{\int d\Omega_0}{\int dS}=(1+z_s)^{-4}\frac{4\pi}{\Acal(w_o, z_s) } ~~, \nonumber \\
\Acal(w_o, z_s) &=&  \int _{\Sigma(w_o,z)} d^2 \xi  \sqrt{\ga} ~~.
\eea
Here $\Acal$ is the proper area of $\Sigma(w_o,z_s)$ computed with the induced metric $\gamma_{ab}$ \footnote{As well known from the Nambu-Goto action in string theory!}.  Exactly the same result follows from the averaging prescription of \cite{GMNV}, which uses an alternative (and equivalent) definition of the surface  
$\Sigma(w_o,z_s)$ through two constraints obeyed by the coordinates. Eq. (\ref{Phitheory}) holds non-perturbatively for any 
space-time and was the starting point of the computation of the average flux presented in \cite{BGMNV2} (see Eq. (6) therein).
It can also be written in an elegant form in which  the flux of an inhomogeneous Universe  is compared to that of a FLRW one:
\bea
\label{avfluxgen}
\langle \Phi/\Phi^{FLRW} \rangle = \left( d_A^{FLRW}\right)^2  ~\frac{4\pi}{\Acal} ~~.
\eea

It is now straightforward to formally express the result in a different gauge by simply changing the GLC gauge coordinates into those of the chosen new gauge. In our specific case we would like to express $\gamma_{ab}$ in terms of PG perturbations and, for our physical application to dark energy, as a function  of the observer's angles
$\ti{\theta}^a_o = \ti{\theta}^a_s$. Taking in (\ref{gammaab}) the coordinates and the metric to be those of the PG we have:
\be
\gamma_{a b}^{ind} = \gamma_{a b} =  \frac{\partial y^{\mu}}{\partial \xi^{a}} \frac{\partial y^{\nu}}{\partial \xi^{b}} g_{\mu \nu}^{PG}(y)
\label{gammaabPG}\, ,
\ee 
where $y^{\mu} = y^{\mu} (w_o, \tau_s(w_o, z_s, \ti{\theta}^a), \ti{\theta}^a)$ define now the surface $\Sigma(w_o, z_s)$ in the PG coordinates $y^{\mu}$.
Unfortunately, it is not trivial to find the explicit form of the above relation. We have found the easiest procedure to consist of: i) expressing the GLC gauge coordinates $x^{\mu}$ in terms of the PG  coordinates and metric; ii)  imposing the condition that the sources lie on $\Sigma$; iii) finally, inverting the second order transformation to express the outcome in terms of the $\ti{\theta}^a$ angles. The first of this three-step process, which is of general interest, is carried out in the following two subsections.


\subsection{The second order transformation for scalar perturbations}

We now generalize to second order the transformation between the GLC gauge and the PG already 
obtained to first order in \cite{BGMNV1}. Before carrying on, let us mention that using the GLC approach means that we are taking many physical effects into account already at the level of the metric. In this approach, the geodesic equations for the observer and light rays are solved non-perturbatively, and their solutions are expressed in terms of the $\tau,w$ coordinates. The outcome is that physical phenomena such as redshift perturbations (RP), redshift space distortions (RSD), Sachs-Wolfe effect (SW), integrated Sachs-Wolfe effect (ISW), peculiar velocities, lensing and others, are manifestly encoded in the metric,
and are derived from a coordinate transformation. This is different from the usual approach, which first takes some perturbed metric, and then solves the geodesic equations order by order to construct physical observables (see, for example, 
\cite{Bonvin,DiDio:2012bu}).

Since, by our assumption, vector and tensor modes 
appear only at second order and, as a consequence, will be decoupled from scalar perturbations,  
we can neglect them momentarily and add them at the end.

Considering only scalar perturbations and using spherical coordinates $(r, \theta^a)=(r, \theta, \phi)$,
the PG metric defined in Eq.(\ref{PGmetricstandard}) can be rewritten as
\be
g_{PG}^{\mu \nu} = a(\eta)^{-2} diag( -1 + 2 \tilde{\Phi}, 1 + 2 \tilde{\Psi}, (1 + 2 \tilde{\Psi}) \gamma_0^{ab} ) ~~,
\ee
where $\ga_0^{ab}=  {\rm diag} \left( r^{-2}, r^{-2} \sin^{-2} \theta \right)$, $\tilde{\Phi} = \psi + \frac{1}{2} \phi^{(2)} - 2 \psi^2$ and $\tilde{\Psi} = \psi + \frac{1}{2} \psi^{(2)} + 2 \psi^2$.
Following \cite{BGMNV1} we compute the GLC gauge (inverse) metric through:
\be
g_{GLC}^{\rho\sigma}(x)=\frac{\partial x^\rho}{\partial y^\mu}
\frac{\partial x^\sigma}{\partial y^\nu} g_{PG}^{\mu\nu}(y) ~~.
\label{EqBetweenGauges}
\ee
where, as before, we indicate  with $y^\mu=(\eta, r, \theta^a)$ the PG coordinates and
with $x^\nu=(\tau, w,\ti{\theta}^a)$ the GLC ones. Let us also introduce the useful (zeroth-order) light-cone variables $\eta_\pm= \eta \pm r$, 
with corresponding partial derivatives:
\beq
\pa_\eta = \pa_+ + \pa_- ~~~,~~~~~ \pa_r = \pa_+ - \pa_- ~~~,~~~~~\pa_\pm= {\pa \over \pa \eta_\pm}={1\over 2} \left( \pa_\eta \pm \pa_r \right) ~~.
\eeq

Using these variables we solve the four differential equations obtained from Eq. (\ref{EqBetweenGauges}) for the components $g_{GLC}^{\tau \tau} = -1$, $g_{GLC}^{ww} = 0$, $g_{GLC}^{wa} = 0$, by imposing the boundary conditions that $i)$ the transformation is non singular around $r=0$, and $ii)$ that the two-dimensional spatial sections $r=$ const are locally parametrized at the observer's position by standard spherical coordinates. 

To this purpose we also introduce the following auxiliary quantities:
\be
P(\eta, r, \theta^a) = \int_{\eta_{in}}^\eta d\eta' \frac{a(\eta')}{a(\eta)} \psi(\eta',r,\theta^a)
\,\,\,\,\,\,\,\,,\,\,\,\,\,\,\, Q(\eta_+, \eta_-, \theta^a) = \int_{\eta_+}^{\eta_-} dx~ \hat{\psi}(\eta_+,x,\theta^a) ~~,
\label{PQ}
\ee
where, hereafter, we use a hat to denote a quantity expressed in terms of $(\eta_+,\eta_-,\theta^a)$ variables, for instance $\hat{\psi}(\eta_+,\eta_-,\theta^a) \equiv \psi(\eta,r,\theta^a)$.

The sought for transformation can then be written, to second order in perturbation theory and with self-explanatory notations, 
as follows:
\bea
\tau &=& \tau^{(0)}+\tau^{(1)}+\tau^{(2)} \nonumber \\
&\equiv& \left( \int_{\eta_{in}}^\eta d\eta' a(\eta') \right) + a(\eta) P(\eta, r, \theta^a) 
+ \int_{\eta_{in}}^\eta d\eta' \frac{a(\eta')}{2} \left[ \phi^{(2)} - \psi^2 + ( \partial_r P )^2 + \gamma_0^{ab} ~ \partial_a P ~ \partial_b P \right] (\eta', r, \theta^a)~,
\label{tau2order} \\
w &=& w^{(0)}+w^{(1)}+w^{(2)} \nonumber \\
&\equiv& \eta_+ + Q(\eta_+, \eta_-, \theta^a) + \frac{1}{4} \int_{\eta_+}^{\eta_-} dx~ \left[ \hat{\psi}^{(2)} + \hat{\phi}^{(2)} + 4 \hat{\psi} \partial_+ Q + \hat{\gamma}_0^{ab} ~ \partial_a Q ~ \partial_b Q \right] (\eta_+, x, \theta^a) ~~,
\label{w2order} \\
\tilde{\theta}^a &=& \tilde{\theta}^{a (0)}+\tilde{\theta}^{a (1)}+\tilde{\theta}^{a (2)} \nonumber \\
&\equiv& \theta^a + \frac12 \int_{\eta_+}^{\eta_-} dx~ \left[ \hat{\gamma}_0^{ab} \partial_b Q \right] (\eta_+,x,\theta^a) 
+ \int_{\eta_+}^{\eta_-} dx~ \left[ \hat{\gamma}_0^{ac} \zeta_c + \hat{\psi} ~ \xi^a + \lambda^a \right] (\eta_+,x,\theta^a) ~~,
\label{thetatilde2orderShort}
\eea
where $\eta_{in}$ represents an early enough time when the perturbation (or better the integrand) was negligible. In other words, all the relevant integrals (i.e. for all scales of interest) are insensitive to the actual value of $\eta_{in}$. Furthermore, we have used the following shorthand notations:
\bea
\zeta_c(\eta_+,x,\theta^a) &=& \frac12 \partial_c w^{(2)} (\eta_+,x,\theta^a) = \frac18 \int_{\eta_+}^x du~ \partial_c \left[ \hat{\psi}^{(2)} + \hat{\phi}^{(2)} + 4 \hat{\psi} ~ \partial_+ Q + \hat{\gamma}_0^{ef} ~ \partial_e Q ~ \partial_f Q \right] (\eta_+,u,\theta^a) ~~,
\label{ExprChi}
\\
\xi^a(\eta_+,x,\theta^a) &=& \partial_+ \tilde{\theta}^{a(1)}(\eta_+,x,\theta^a) +2\partial_x\tilde{\theta}^{a(1)}(\eta_+,x,\theta^a) ~~ \nonumber \\
&=& \partial_+ \left( \frac12 \int_{\eta_+}^x du~ [\hat{\gamma}_0^{ac} \partial_c Q] (\eta_+,u,\theta^a) \right) + [\gamma_0^{ac} \partial_c Q] (\eta_+,x,\theta^a) ~~,
\label{ExprXi}
\\
\lambda^a(\eta_+,x,\theta^a) &=& \partial_x \tilde{\theta}^{d(1)} (\eta_+,x,\theta^a)\left(\partial_d \tilde{\theta}^{a(1)} (\eta_+,x,\theta^a)-\delta_d^a  \partial_+ Q (\eta_+,x,\theta^a)\right) ~~ \nonumber \\
&=& \frac14 [ \hat{\gamma}_0^{dc} ~ \partial_c Q ] (\eta_+,x,\theta^a) \left( \int_{\eta_+}^x du ~ \partial_d \left[ \hat{\gamma}_0^{ae} \partial_e Q \right] (\eta_+,u,\theta^a) \right) - \frac12 \left[ \partial_+ Q ~ \hat{\gamma}_0^{ab} \partial_b Q \right] (\eta_+,x,\theta^a) ~~. 
\label{ExprLamb}
\eea
Let us now compute the various non-trivial entries of the GLC metric.
Using $\Ups^{-1} = - \partial_{\mu} w ~\partial_{\nu} \tau ~g_{PG}^{\mu \nu}$, we obtain
\be
\Ups^{-1} = \frac{1}{a(\eta)} \left( 1 + \epsilon^{(1)} + \epsilon^{(2)} \right) ~~.
\label{Ups}
\ee
In terms of quantities implicitly defined in Eqs. \rref{tau2order}-(\ref{w2order}) we find:
\bea
\epsilon^{(1)} &=& \partial_+ Q - \partial_r P ~~,
\label{eps1} \\
\epsilon^{(2)} &=& \partial_{\eta}w^{(2)} + \frac{1}{a}(\partial_\eta - \partial_r) \tau^{(2)} - \psi \partial_{\eta}Q - \phi^{(2)} + 2\psi^2 - \partial_r P \partial_r Q - 2 \psi \partial_r P - \gamma^{ab}_0 \partial_a P \partial_b Q ~~.
\label{eps2}
\eea
The full explicit expression for $\epsilon^{(2)}$ can be then written as follows:
\bea
\epsilon^{(2)} &=&  \frac{1}{4} \left( \psi^{(2)} - \phi^{(2)} \right) + \frac{\psi^2}{2}
+ \frac12 (\partial_r P)^2  - (\psi + \partial_+ Q) \cdot \partial_r P +  \frac14 \gamma_0^{ab}\left( 2\partial_a P \cdot \partial_b P + \partial_a Q \cdot \partial_b Q  - 4  \partial_a Q \cdot \partial_b P \right) \nonumber \\
&+& \frac14~ \partial_+ \int_{\eta_+}^{\eta_-} dx~  \left[ \hat{\psi}^{(2)} + \hat{\phi}^{(2)} + 4 \hat{\psi} ~ \partial_+ Q + \hat{\gamma}_0^{ab} \partial_a Q \cdot \partial_b Q \right] (\eta_+,x,\theta^a) \nonumber \\
&-& \frac12 \int_{\eta_{in}}^\eta d\eta'~ \frac{a(\eta')}{a(\eta)} \partial_r \left[ \phi^{(2)} - \psi^2 + \left( \partial_r P \right)^2 + \gamma_0^{ab} \partial_a P \cdot \partial_b P \right] (\eta',r,\theta^a) ~~,
\label{eps2} 
\eea
where the variables $(\eta,r, \theta^a)$ have been omitted for the sake of conciseness.
The computation of the GLC functions $U^a$ gives:
\bea
U^a &=&-\left\{-\partial_{\eta}\tilde{\theta}^{a (1)}+\frac{1}{a}\gamma_0^{ab}\partial_b \tau^{(1)}-\partial_{\eta}\tilde{\theta}^{a (2)} +\frac{1}{a}\gamma^{ab}_0\partial_b \tau^{(2)} + \frac{1}{a} \partial_r \tau^{(1)} \partial_r \tilde{\theta}^{a(1)} \right. \cr
& +& \left. \psi \left(\partial_\eta \tilde{\theta}^{a (1)}+\frac{2}{a}\gamma^{ab}_0 \partial_b \tau^{(1)} \right)+\frac{1}{a}\gamma_0^{cd}\partial_c \tau^{(1)} \partial_d \tilde{\theta}^{a (1)}-\epsilon^{(1)} \left(-\partial_{\eta} \tilde{\theta}^{a (1)}+\frac{1}{a}\gamma^{ab}_0 \partial_b \tau^{(1)}\right)\right\}~,
\label{Ua}
\eea
where $\tau^{(1),(2)}$, $\tilde \theta^{a (1),(2)}$ are implicitly defined in the coordinate transformations (\ref{tau2order}), and (\ref{thetatilde2orderShort}). $U^a$ is a measure of anisotropy of space-time in GLC coordinates. 
Substituting in \rref{Ua} the explicit values of $\tau^{(1),(2)}$ and $\tilde{\theta}^{a (1),(2)}$, 
we obtain the following explicit expressions for $U^a$:
\bea
U^a &=& \Bigg\{ \gamma_0^{ab} \left( \frac12 \partial_b Q - \partial_b P \right) + \partial_+ \left( \frac12 \int_{\eta_+}^{\eta_-} dx ~ \left[ \hat{\gamma}_0^{ab} ~ \partial_b Q \right] (\eta_+,x,\theta^a) \right) \Bigg\} \nonumber \\
&+& \Bigg\{ - \left( \psi + \partial_+ Q \right) \partial_+ \left( \frac12  \int_{\eta_+}^{\eta_-} dx ~ \left[ \hat{\gamma}_0^{ab} ~ \partial_b Q \right] (\eta_+,x,\theta^a) \right) + \left( - \psi + 2 \partial_r P - \partial_+ Q \right) \frac12 \gamma_0^{ab} ~ \partial_b Q \nonumber \\
& & ~~~~~ - 2 \psi \gamma_0^{ab} ~ \partial_b P - \frac12 \gamma_0^{cd} ~ \partial_c P \int_{\eta_+}^{\eta_-} dx~ \partial_d \left[ \hat{\gamma}_0^{ab} ~ \partial_b Q \right] (\eta_+,x,\theta^a) + ( \partial_+ Q - \partial_r P ) \gamma_0^{ab} \partial_b P \nonumber \\
& & ~~~~~ - \frac12 \gamma_0^{ab} \int_{\eta_{in}}^{\eta} d\eta' \frac{a(\eta')}{a(\eta)} \partial_b \left[ \phi^{(2)} - \psi^2 + (\partial_r P)^2 + \gamma_0^{cd} \partial_c P \partial_d P \right](\eta',r,\theta^a) \nonumber \\
& & ~~~~~ + (\partial_+ + \partial_-) \int_{\eta_+}^{\eta_-} dx~ \left[ \hat{\gamma}_0^{ac} \zeta_c + \hat{\psi} ~ \xi^a + \lambda^a \right] (\eta_+,x,\theta^a) \Bigg\} ~~.
\eea
Finally, starting from $\gamma^{a b} = \frac{\partial \tilde{\theta}^a}{\partial y^\mu} \frac{\partial \tilde{\theta}^b}{\partial y^\nu} g_{PG}^{\mu\nu}(y)$, we find:
\bea
a(\eta)^2 \gamma^{ab} &=& \gamma_0^{ab} \left(1 +  2 \psi\right) +\left[\gamma_0^{a c} \partial_c \tilde{\theta}^{b (1)}+ (a\leftrightarrow b) \right]+ \gamma_0^{ab}
\left(\psi^{(2)} + 4 \psi^2 \right)-\partial_\eta \tilde{\theta}^{a (1)}\partial_\eta \tilde{\theta}^{b (1)}+\partial_r \tilde{\theta}^{a (1)}\partial_r \tilde{\theta}^{b (1)} \nonumber \\
&+& 2 \psi \left[\gamma_0^{a c} \partial_c \tilde{\theta}^{b (1)}+ (a\leftrightarrow b) \right]+\gamma_0^{c d} \partial_c \tilde{\theta}^{a (1)}
 \partial_d \tilde{\theta}^{b (1)}+\left[\gamma_0^{a c} \partial_c \tilde{\theta}^{b (2)}+ (a\leftrightarrow b) \right] \, .
\label{gammaab1}
\eea
More explicitly, in terms of the quantities defined in (\ref{PQ}) and in (\ref{ExprChi})-(\ref{ExprLamb}):
\bea
a(\eta)^2 \gamma^{ab} &=& \gamma_0^{ab} \left(1 +  2 \psi\right) + \frac12 \bigg\{\gamma_0^{ad} \int_{\eta_+}^{\eta_-} dx~ \partial_d \left[ \hat{\gamma}_0^{bc} \partial_c Q \right] (\eta_+,x,\theta^a) + (a\leftrightarrow b) \bigg\} \nonumber \\
&+& \left(\psi^{(2)} + 4 \psi^2 \right) \gamma_0^{ab} - \bigg\{ \gamma_0^{ac} \partial_c Q ~ \partial_+ \left(\frac12 \int_{\eta_+}^{\eta_-} dx~  \left[ \hat{\gamma}_0^{bd} ~ \partial_d Q \right] (\eta_+,x,\theta^a) \right) + (a\leftrightarrow b) \bigg\} \nonumber \\
&+& \psi \bigg\{ \gamma_0^{ad} \int_{\eta_+}^{\eta_-} dx~ \partial_d \left[ \hat{\gamma}_0^{bc} \partial_c Q \right] (\eta_+,x,\theta^a) + ( a \leftrightarrow b ) \bigg\} \nonumber \\
&+& \frac{1}{4} \gamma_0^{cd} \bigg(\int_{\eta_+}^{\eta_-} dx~ \partial_c \left[ \hat{\gamma}_0^{ae} \partial_e Q \right] (\eta_+,x,\theta^a) \bigg) \bigg( \int_{\eta_+}^{\eta_-} d\bar{x}~ \partial_d \left[ \hat{\gamma}_0^{bf} \partial_f Q \right] (\eta_+,\bar{x},\theta^a) \bigg) \nonumber \\
&+& \bigg\{ \gamma_0^{ac} \int_{\eta_+}^{\eta_-} dx~ \partial_c \left[ \hat{\gamma}_0^{bd} ~ \zeta_d + \hat{\psi} ~ \xi^b + \lambda^b \right] (\eta_+,x,\theta^a) + ( a \leftrightarrow b ) \bigg\} ~~.
\label{gammaab2}
\eea


\subsection{The second order transformation for vector and tensor perturbations}

As already mentioned we can add the contributions of the tensor and vector perturbations by considering them separately. 
Using spherical coordinates $(r, \theta^a)=(r, \theta, \phi)$,
the tensor and vector part of the PG metric defined in Eq.(\ref{PGmetricstandard}) can be rewritten as:
\beq
ds_{PG}^2 = a^2(\eta) \left[ -d\eta^2 + 2 v_i d\eta d x^i + [(\gamma_0)_{ij} + \chi_{ij}] d x^i d x^j \right]\, ,
\eeq
corresponding to:
\beq
g_{PG}^{\mu\nu}(\eta,r,\theta^a) = a^{-2}(\eta)
\left(
\begin{array}{cc}
-1 & v^i \\
v^j & \gamma_0^{ij} - \chi^{ij} \\
\end{array}
\right) ~,
\eeq
where  $\gamma_0^{ij} = diag (1, r^{-2}, r^{-2} (\sin\theta)^{-2} )$ is the (inverse) flat 3-metric. 
Here $v^i$ and $\chi^{ij}$  are the vector and 
tensor perturbation in spherical coordinates equivalent to the more standard definition
$\omega^i$ and $h^{ij}$ used in cartesian coordinates. 
They satisfy  $\nabla_i v^i=\nabla_i \chi^{i j}=0$ and $(\gamma_0)_{ij} \chi^{ij}=0$ with $\nabla_i$ the flat covariant derivative in spherical coordinates.

Proceeding as for the scalar part of the metric, we first note that $\tau$ is not affected by vector and tensor perturbations,
\beq
\tau = \int_{\eta_{in}}^\eta d\eta' a(\eta') ~~, 
\eeq
while the light-cone coordinate $w$ is:
\beq
\label{wVT}
w = \eta_+  + Q^{(\alpha)}(\eta_+,\eta_-,\theta^a) ~~,
\eeq
where
\beq
\label{Qalpha}
Q^{(\alpha)}(\eta_+,\eta_-,\theta^a) = \int_{\eta^+}^{\eta^-} dx ~ \hat{\alpha}^r (\eta^+,x,\theta^a)~~~~~ \mbox{ with } ~~~~~ \alpha^r \equiv \frac{v^r}{2} - \frac{\chi^{rr}}{4} ~~.
\eeq
As before, hats mean that $(\eta_+,\eta_-,\theta^a)$-coordinates are used.
Finally, we find:
\be
\tilde{\theta}^a = \tilde{\theta}^{a (0)}+\tilde {\theta}^{a (2)}=\theta^a + \frac{1}{2} \int_{\eta_+}^{\eta-} dx \left( \hat{v}^a(\eta_+,x,\theta^a) - \hat{\chi}^{ra}(\eta_+,x,\theta^a) + \hat{\gamma}^{ab}_0(\eta_+,x,\theta^a) \int_{\eta_+}^x dy~ \partial_b \hat{\alpha}^r (\eta_+,y,\theta^a) \right) ~.
\ee
Following the same steps as for scalar perturbations, we then compute the non-trivial entries of the GLC metric:
\bea
a(\eta) \Ups^{-1} &=& 1 - \frac{1}{2} v^r - \frac{1}{4} \chi^{rr} + \partial_+  \int_{\eta_+}^{\eta_-} dx~  \hat{\alpha}(\eta_+,x,\theta^a) ~~, \label{UpsVT} \\
a(\eta)^2 \gamma^{ab} &=& \gamma_0^{ab} - \chi^{ab} \\
& +& \left[ \frac{\gamma_0^{ac}}{2} \int_{\eta_+}^{\eta_-} dx~ \partial_c \left( \hat{v}^b(\eta_+,x,\theta^a) - \hat{\chi}^{rb}(\eta_+,x,\theta^a) + \hat{\gamma}_0^{bd} (\eta_+,x,\theta^a) \int_{\eta_+}^{x} dy~ \partial_d \hat{\alpha}^r (\eta_+,y,\theta^a) \right) + (a \leftrightarrow b) \right] ~~, \label{gammaVT} \nonumber \\
U^a &=& - v^a + \partial_{\eta}\tilde {\theta}^{a (2)} = -v^a+ \partial_{-}\tilde {\theta}^{a (2)} + \partial_{+}\tilde {\theta}^{a (2)}
= \frac{1}{2} \left( - v^a - \chi^{ra} + \gamma^{ab}_0 \int_{\eta_+}^{\eta_-} dx ~ \partial_b \hat{\alpha}^r (\eta_+,x,\theta^a) \right) \nonumber \\
&+& \frac{1}{2} \partial_+ \left( \int_{\eta_+}^{\eta-} dx~ \left[\hat{v}^a(\eta_+,x,\theta^a) - \hat{\chi}^{ra}(\eta_+,x,\theta^a) + \hat{\gamma}^{ab}_0(\eta_+,x,\theta^a) \int_{\eta_+}^x dy~ \partial_b \hat{\alpha}^r (\eta_+,y,\theta^a) \right] \right) ~~,
\eea
where, again, the variables $(\eta,r,\theta^a)$ have been omitted for the sake of  conciseness. Of course, these vector and tensor corrections to the FLRW metric have to be added to the scalar ones of the previous subsection.


\section{Detailed expression for $d_L( z, \theta^a)$}
\label{Sec4}
\setcounter{equation}{0}


\subsection{The scalar  contribution}

We now apply the above coordinate transformations  to find the final expression of the luminosity distance in terms of  perturbations in the PG, of the observed redshift, and of the observer's angular coordinates beginning, once more, with the scalar contribution.
From Eqs.(\ref{dLdA}) and (\ref{Eq_dA}) we have:
\be
d_L=(1+z_s)^2~\gamma^{1/4}~(\sin \tilde{\theta})^{-1/2}   ~~.
\label{ExactdL}
\ee
Let us start with $\ga$. For a source emitting light at time $\eta_s$ and radial distance $r_s$ we obtain from Eq. (\ref{gammaab1}): 
\bea
\gamma^{-1} \equiv \det \gamma^{ab} &=& 
(a_s^2r_s^2\sin\theta)^{-2} \left\{1 + 4 \psi_s + 2 \partial_a \tilde{\theta}^{a (1)} +2 \psi_s^{(2)}
+12 \psi_s^2+ 2 \partial_a \tilde{\theta}^{a (2)}
-4 \gamma_{0 a b} \partial_+ \tilde{\theta}^{a (1)}\partial_- \tilde{\theta}^{b (1)}
\right. \nonumber \\
& & ~~~~~ ~~~~~ ~~~~ ~~~~ ~~~ \left.
+ ~ 8 \psi_s \partial_a \tilde{\theta}^{a (1)}+2 \partial_a \tilde{\theta}^{a (1)}
\partial_b \tilde{\theta}^{b (1)}
-\partial_a \tilde{\theta}^{b (1)}\partial_b \tilde{\theta}^{a (1)}
\right\} ~~.
\label{det}
\eea
We also need the expression for 
$\sin \tilde{\theta}$ up to second order in perturbation theory. This is easily given as:
\be
\sin \tilde{\theta}= \sin \theta \left[1+\cot \theta \left(  \tilde{\theta}^{ (1)}
+\tilde{\theta}^{ (2)}\right)-\frac{1}{2}\left(\tilde{\theta}^{ (1)}\right)^2\right] ~~.
\label{sinThetaTilde}
\ee
Using Eqs.(\ref{det}) and (\ref{sinThetaTilde}), Eq.(\ref{ExactdL})  yields:
\bea
d_L &=& (1+z_s)^2 (a_s r_s) \left\{1-\psi_s - J_2 - \frac{1}{2}\psi_s^{(2)} - \frac{1}{2} \psi_s^2 - K_2 + \psi_s J_2 + \frac{1}{2} (J_2)^2 + \frac{1}{4 \sin^2 \theta}\left(\tilde{\theta}^{ (1)}\right)^2 \right. \nonumber \\
& & ~~~~~ ~~~~~ ~~~~~ ~~~~ ~~~~ \left. + ~(\gamma_0)_{a b} \partial_+ \tilde{\theta}^{a (1)}\partial_- \tilde{\theta}^{b (1)}
+ \frac{1}{4}\partial_a \tilde{\theta}^{b (1)}\partial_b \tilde{\theta}^{a (1)}
\right\} ~~,
\label{d_L}
\eea
where:
\be
J_2 = \frac{1}{2}\left[\cot \theta ~ \tilde{\theta}^{(1)}+\partial_a \tilde{\theta}^{a (1)}\right] \equiv \frac12 \nabla_a \tilde{\theta}^{a (1)}
\,\,\,\,\,\,\,\,\,,\,\,\,\,\,\,\,\,\,
K_2 = \frac{1}{2}\left[\cot \theta ~ \tilde{\theta}^{(2)}+\partial_a \tilde{\theta}^{a (2)}\right] \equiv \frac12 \nabla_a \tilde{\theta}^{a (2)}
~~.
\ee
All the above quantities are evaluated at the source (apart from $\psi_s$, we neglect the suffix $s$ for simplicity).

At this point, for the explicit expression of the luminosity distance $d_L$ at constant redshift, we need the  first and second-order expansion of the factor $a_sr_s\equiv a(\eta_s) r_s$ appearing in Eq. (\ref{d_L}). To this purpose, we start from the explicit expression for the redshift parameter $z_s$ (see  Eq.(\ref{redshift})), 
considered as a constant parameter localizing, together with the $w = w_o$ condition, the source on $\Sigma(w_o, z_s)$. We then look for approximate solutions for $\eta_s= \eta_s(z_s, \theta^a)$ and $r_s= r_s(z_s, \theta^a)$. 

Let us first define the zero-order solution ${\eta}_s^{(0)}$  through the (exact) relation:
\be
\label{OnePlusZHom}
\frac{a({\eta}_s^{(0)})}{a_o} = \frac{1}{1+z_s} ~~,
\ee
where $a_o \equiv a(\eta_o)$. 
Inserting now the result (\ref{Ups}) into  Eq. (\ref{redshift}) and expanding $a(\eta_s)$ and $\Ups^{-1}$ with respect to the background solutions $\eta_s^{(0)}$ and $r_s^{(0)}$ 
(where we define $\eta_s=\eta_s^{(0)} + \eta_s^{(1)} + \eta_s^{(2)}$ and
$r_s=r_s^{(0)} + r_s^{(1)} + r_s^{(2)}$), we obtain:
\bea
\label{OnePlusZInom}
\frac{1}{1 + z_s} &=& \frac{a(\eta_s^{(0)})}{a(\eta_o)} \Bigg\{1 \frac{}{} + \left[ {\mathcal H}_s \eta_s^{(1)} + \epsilon_o^{(1)} - \epsilon_s^{(1)} \right] 
+ \bigg[ {\mathcal H}_s \eta_s^{(2)} + ({\mathcal H}_s' + {\mathcal H}_s^2)\frac{(\eta_s^{(1)})^2}{2} + \epsilon_o^{(2)} - \epsilon_s^{(2)} - \epsilon_s^{(1\rightarrow 2)}
\nonumber \\
& & 
\,\,\,\,\,\,\,\,\,\,\,\,\,\,\,\,\,\,\,\,\,\,\,\,\,\,+ (\epsilon_s^{(1)})^2 - \epsilon_o^{(1)} \epsilon_s^{(1)} 
+ {\mathcal H}_s \eta_s^{(1)} (\epsilon_o^{(1)} - \epsilon_s^{(1)}) \bigg] \Bigg\} ~~, 
\label{1over1plusz}
\eea
where ${\mathcal H}_s=\frac{a'(\eta_s^{(0)})}{a(\eta_s^{(0)})}$ and
\be
\epsilon_o^{(1)}=\epsilon^{(1)}(\eta_o, 0, \theta^a)\,\,\,\,\,,\,\,\,\,\,
\epsilon_s^{(1)}=\epsilon^{(1)}(\eta_s^{(0)}, r_s^{(0)}, \theta^a)\,\,\,\,\,,\,\,\,\,\,
\epsilon_o^{(2)}=\epsilon^{(2)}(\eta_o, 0, \theta^a)\,\,\,\,\,,\,\,\,\,\,
\epsilon_s^{(2)}=\epsilon^{(2)}(\eta_s^{(0)}, r_s^{(0)}, \theta^a) ~~,
\label{epsis}
\ee
\be
\epsilon_s^{(1\rightarrow 2)} = \left[\partial_\eta \epsilon^{(1)}\right](\eta_s^{(0)}, r_s^{(0)}, \theta^a) ~ \eta_s^{(1)} + \left[\partial_r \epsilon^{(1)}\right](\eta_s^{(0)}, r_s^{(0)}, \theta^a) ~ r_s^{(1)} ~~.
\label{eps12}
\ee

Similarly, in order to compute $ r_s^{(1)}$ and $ r_s^{(2)}$, we need to expand the $w = w_o$ constraint by writing:
\be
\label{ws1,2}
w_o = \left\{ \eta_s^{(0)} + r_s^{(0)} \right\} + \left\{ \eta_s^{(1)} + r_s^{(1)} + w_s^{(1)} \right\} + \left\{ \eta_s^{(2)} + r_s^{(2)} + 
w_s^{(2)} + w_s^{(1 \rightarrow 2)} \right\}
\ee
where
\be
\label{ws12a}
w_s^{(1)} = w^{(1)}(\eta_s^{(0)}, r_s^{(0)}, \theta^a) \,\,\,\,\,,\,\,\,\,\, w_s^{(2)} = w^{(2)}(\eta_s^{(0)}, r_s^{(0)}, \theta^a) ~~,
\ee
\be
\label{ws12}
w_s^{(1 \rightarrow 2)} = \left[\partial_\eta w^{(1)}\right](\eta_s^{(0)}, r_s^{(0)}, \theta^a) ~ \eta_s^{(1)} + \left[\partial_r w^{(1)}\right] (\eta_s^{(0)}, r_s^{(0)}, \theta^a) ~ r_s^{(1)} ~~.
\ee
The additional terms $\epsilon_s^{(1 \rightarrow 2)}$, $w_s^{(1 \rightarrow 2)}$ appearing in the above equations stand for the second order contributions coming from Taylor expanding $\epsilon_s^{(1)}$, $w_s^{(1)}$,  around the background source 
position~\footnote{There is no equivalent contribution at the observer position as $\eta_o$ and $r_o = 0$ are fixed quantities, with no perturbative corrections.}. 
More precisely,
they originate from the fact that, at first order, quantities that are already first order are integrated along the unperturbed line of sight, while, at second order,  first order terms have to be integrated along the \textit{perturbed} line of sight.

From Eq.(\ref{eps1}) we obtain~\footnote{In this paper we use $Q_s$ instead of the quantity  $\Psi_{av}$  introduced in \cite{BGMNV2}, the two   are directly related by $Q_s \equiv - 2 \Delta \eta \Psi_{av}$.}:
\be
\epsilon_s^{(1)} - \epsilon_o^{(1)} = J 
\ee 
with
\be
J ~~ \equiv ~ ([\partial_+ Q]_s - [\partial_+ Q]_o) - ([\partial_r P]_s - [\partial_r P]_o) ~~,
\ee
and where, for example, the term $[ \partial_r P ]_s$ denotes the expression of $\partial_r P$ with $\eta$ and $r$ replaced by $\eta_s^{(0)}$ and $r_s^{(0)}$, we also remark that $[\partial_+ Q]_o = - \psi_o$. 
Then, from Eqs.(\ref{1over1plusz}), (\ref{ws1,2}), recalling (\ref{w2order}), we compute:
\bea
\eta_s^{(1)}=\frac{J}{\Hcal_s} ~~~~~,~~~~~ r_s^{(0)} = \eta_o - \eta_s^{(0)} \equiv \Delta \eta ~~~~~,~~~~~ r_s^{(1)} = - Q_s - \frac{J}{{\mathcal H}_s} ~~~.
\eea
These expressions are in accordance with our previous work \cite{BGMNV1}.
We wish to note, already at this point, that the $\epsilon$ terms correspond to redshift perturbations (RP). The first order term, $\epsilon^{(1)}$, gives rise to the Doppler effect due to the peculiar velocities $\partial_rP$, and the SW and ISW effects  are combined together in $\partial_+ Q_s-\partial_+ Q_o$.
Let us, in fact, recall that $\partial_r P$ can be rewritten as \cite{BGMNV1}
\be
\label{doppler1}
\partial_r P =    \vec{v} \cdot  \hat {n}, 
\ee
where $\hat n$ is the unit tangent vector along the null geodesic connecting source and observer, and where 
\be
\vec{v}=-\int_{\eta_{in}}^{\eta}d\eta'\frac{a(\eta')}{a(\eta)}\vec{\nabla} \Psi(\eta', r, \theta^a)
\ee
are the  ``peculiar velocities" associated to a  geodesic configuration perturbed up to first order in the PG.

Let us now move to the 
second order quantities appearing in (\ref{epsis}, \ref{eps12}) and (\ref{ws12a}, \ref{ws12}).
We simply have, from Eq.(\ref{eps2}),
\bea
\epsilon_s^{(2)} - \epsilon_o^{(2)} &=& - \frac{1}{4} \left( \phi_s^{(2)} - \phi_o^{(2)} \right) + \frac{1}{4} \left( \psi_s^{(2)} - \psi_o^{(2)} \right) + \frac12 \left( \psi_s^2 - \psi_o^2 \right) + \frac12 ([\partial_r P]_s)^2 - \frac12 ([\partial_r P]_o)^2 - \left( \psi_s + [\partial_+ Q]_s \right) \cdot [\partial_r P]_s
\nonumber \\
& &
+ \frac14 (\gamma_0^{ab})_s \left( 2 \partial_a P_s \cdot \partial_b P_s + \partial_a Q_s \cdot \partial_b Q_s - 4 \partial_a Q_s \cdot \partial_b P_s \right) - \frac12 \lim_{r\rightarrow 0} \left[\gamma_{0}^{ab} \partial_a P \cdot \partial_b P \right] \nonumber \\
& & 
- \frac12 \int_{\eta_{in}}^{\eta_s^{(0)}} d\eta' \frac{a(\eta')}{a(\eta_s^{(0)})} \partial_r \left[ \phi^{(2)} - \psi^2 + (\partial_r P)^2 + \gamma_0^{ab} \partial_a P \cdot \partial_b P \right] (\eta', \Delta \eta, \theta^a) \nonumber \\
& & 
+ \frac12 \int_{\eta_{in}}^{\eta_o} d\eta' \frac{a(\eta')}{a(\eta_o)}  \partial_r \left[ \phi^{(2)} - \psi^2 + (\partial_r P)^2 + \gamma_0^{ab} \partial_a P \cdot \partial_b P \right] (\eta', 0, \theta^a) \nonumber \\
& & 
+ \frac14 \int_{\eta_s^{(0)+}}^{\eta_s^{(0)-}} dx~ \partial_+ \left[ \hat{\phi}^{(2)} + \hat{\psi}^{(2)} + 4 \hat{\psi} ~ \partial_+ Q + \hat{\gamma}_{0}^{ab} \cdot \partial_a Q \cdot \partial_b Q \right] (\eta_s^{(0)+}, x, \theta^a) ~~, 
\eea
while
\be
\epsilon_s^{(1 \rightarrow 2)} = Q_s \left\{ - [\partial_+^2 Q]_s + [\partial_+ \hat{\psi}]_s + [\partial_r^2 P]_s \right\} + \frac{J}{{\mathcal H}_s} \left\{ [\partial_\eta \psi]_s + {\mathcal H}_s [\partial_r P]_s + [\partial_r^2 P]_s \right\} \,.
\ee
We then have:
\bea
w_s^{(1 \rightarrow 2)} &=& \frac{2}{{\mathcal H}_s} \psi_s J + Q_s \left( \psi_s - [\partial_+ Q]_s \right) ~~.
\eea
Using  (\ref{OnePlusZInom}), (\ref{ws1,2}), and recalling (\ref{w2order}), we can now calculate $\eta_s$ and $r_s$ to second order obtaining:
\bea
\eta_s^{(2)}&=&-\frac{1}{\Hcal_s}\left\{({\mathcal H}_s' + {\mathcal H}_s^2)\frac{(\eta_s^{(1)})^2}{2} + \epsilon_o^{(2)} - \epsilon_s^{(2)} - \epsilon_s^{(1\rightarrow 2)} 
+ (\epsilon_s^{(1)})^2 - \epsilon_o^{(1)} \epsilon_s^{(1)} 
+ {\mathcal H}_s \eta_s^{(1)} (\epsilon_o^{(1)} - \epsilon_s^{(1)})\right\}\nonumber \\
&=&-\frac{1}{\Hcal_s}\left\{\frac{{\mathcal H}_s' + {\mathcal H}_s^2}{\Hcal_s^2}\frac{J^2}{2} + \epsilon_o^{(2)} - \epsilon_s^{(2)} - \epsilon_s^{(1\rightarrow 2)} 
+ \epsilon_s^{(1)}J 
- J^2\right\} ~~,
\eea 
and
\bea
r_s^{(2)}&=&-\left( \eta_s^{(2)} + w^{(2)}_s + w_s^{(1 \rightarrow 2)} \right) \nonumber \\
 &=& - \frac{1}{{\mathcal H}_s} \left( \epsilon_s^{(2)} - \epsilon_o^{(2)} + \epsilon_s^{(1 \rightarrow 2)} \right) - \frac{J}{{\mathcal H}_s} \left( 2 \psi_s + \psi_o + [\partial_r P]_o \right) + \frac{{\mathcal H}_s^2 + {\mathcal H}_s'}{2{\mathcal H}_s^3} J^2 + \left( - \psi_s + [ \partial_+ Q ]_s \right) Q_s \nonumber \\
& & - \frac14 \int_{\eta_s^{(0)+}}^{\eta_s^{(0)-}} dx~ \left[ \hat{\phi}^{(2)} + \hat{\psi}^{(2)} + 4 \hat{\psi} ~ \partial_+ Q + \hat{\gamma}_0^{ab} ~ \partial_a Q ~ \partial_b Q \right] (\eta_s^{(0)+},x,\theta^a) ~~.
\eea

In the second order terms we have the expected couplings between first order terms as well as the (also expected) genuine second order SW and ISW effects such as  $(\psi_s^{(2)} - \psi_o^{(2)})$ and $\int dx \partial_{+}\hat{\psi}^{(2)}$. However, at second order, new effects come into play: most notably the tangential peculiar velocity $\partial_aP$, the tangential variation of the photon path $\partial_aQ$, and a RSD due to the peculiar acceleration $\partial^2P$. The somewhat surprising appearance of tangential derivatives in $\eta_s^{(2)}$ and $r_s^{(2)}$ is simply a reflection of working on a fixed-$z$ surface. As a consequence, redshift perturbations originating from those of $\tau$, eq. (\ref{tau2order}),  feed back on $\eta_{s}, r_s$ and, eventually, on $d_L(z)$ .

To conclude, combining these results, we  obtain:
\bea
\frac{a(\eta_s) r_s}{a(\eta_s^{(0)}) \Delta\eta} &=&
1 + \left\{ \Xi_s J - \frac{Q_s}{\Delta \eta} \right\} +
\Bigg\{ \Xi_s \left( \epsilon_s^{(2)} - \epsilon_o^{(2)} + \epsilon_s^{(1 \rightarrow 2)} \right) - \frac{1}{{\mathcal H}_s \Delta \eta} \left( 1 - \frac{{\mathcal H}_s'}{{\mathcal H}_s^2} \right) \frac{J^2}{2} - \frac{2}{{\mathcal H}_s \Delta \eta} \psi_s J \nonumber \\
& & ~ + \Xi_s \left( \psi_o + [\partial_r P]_o \right) J + \left( - \psi_o - \psi_s + [\partial_r P]_s - [\partial_r P]_o \right) \frac{Q_s}{\Delta \eta} \nonumber \\
& & ~ - \frac{1}{4 \Delta\eta} \int_{\eta_s^{(0)+}}^{\eta_s^{(0)-}} dx~ \left[ \hat{\phi}^{(2)} + \hat{\psi}^{(2)} + 4 \hat{\psi} ~ \partial_+ Q + \hat{\gamma}_{0}^{ab} ~ \partial_a Q ~ \partial_b Q \right] (\eta_s^{(0)+},x,\theta^a) \Bigg\} ~~, 
\label{asrs}
\eea
where
\be
\Xi_s \equiv 1 - \frac{1}{\Hcal_s \Delta \eta} ~~.
\ee

Let us also note that in Eq.(\ref{d_L}) there are two other  first order terms that have to be Taylor expanded up to second order around the background solution connected to the observed redshift $z_s$, 
i.e. $\psi_s$ and $J_2$.
We find:
\be
\psi_s =  \psi_s^{(1)}  + \psi_s^{(1 \rightarrow 2)} = \psi(\eta_s^{(0)}, \Delta \eta, \theta^a)+\frac{J}{{\mathcal H}_s} 
\left[ \partial_\eta \psi - \partial_r \psi \right] (\eta_s^{(0)}, \Delta \eta, \theta^a) - Q_s [\partial_r \psi](\eta_s^{(0)}, \Delta \eta, \theta^a) ~~,
\label{ExpPsis}
\ee
\bea
J_2 &=& J_2^{(1)} +  J_2^{(1 \rightarrow 2)} = \frac{1}{\Delta\eta} \int_{\eta_s^{(0)}}^{\eta_o} d \eta' \,\frac {\eta' - \eta_s^{(0)}}{\eta_o - \eta'} \Delta_2 \psi(\eta', \eta_o-\eta', \theta^a) - \left( \frac{J}{{\mathcal H}_s}+ \frac{Q_s}{2} \right) \frac{1}{\Delta\eta^2} \int_{\eta_s^{(0)}}^{\eta_o} d \eta' \Delta_2 \psi(\eta', \eta_o-\eta', \theta^a) \nonumber \\
& & ~~~~~ ~~~~~ ~~~~~~~ ~~~~~~~ ~~~~~~~ - Q_s ~ \partial_+ \left(\int_{\eta_s^{(0)+}}^{\eta_s^{(0)-}} dx \frac{1}{(\eta_s^{(0)+}-x)^2} \int_{\eta_s^{(0)+}}^x d y ~ \Delta_2 \hat{\psi}(\eta_s^{(0)+}, y, \theta^a) \right) ~~, 
\label{ExpJ2}
\eea
where we have used the 2-dimensional Laplacian  $\Delta_2 \equiv \partial^2_{\theta} + \cot \theta\, \partial_{\theta} + (\sin \theta)^{-2} \partial^2_{\phi}$.

Collecting all the results obtained up to now, and inserting them in Eq.(\ref{d_L}), we  write our final result on the effect of scalar perturbations in the following concise form:
\beq
\frac{d_L(z_s, \theta^a)}{(1+z_s)a_o \Delta \eta}
= {d_L(z_s, \theta^a)\over d_L^{FLRW}(z_s)} =  1 + \delta_S^{(1)}(z_s, \theta^a) + \delta_S^{(2)}(z_s, \theta^a) ~~,
\eeq
where:
\bea
\delta_S^{(1)}(z_s, \theta^a) &=& \Xi_s J - \frac{Q_s}{\Delta \eta} - \psi_s^{(1)} - J_2^{(1)} ~~, \nonumber \\
\delta_S^{(2)}(z_s, \theta^a)   &=& - \left(\Xi_s J - \frac{Q_s}{\Delta \eta} \right) \left( \psi_s^{(1)}  + J_2^{(1)} \right)  - \psi_s^{(1 \rightarrow 2)} - J_2^{(1 \rightarrow 2)} + X^{(2)} + Y^{(2)} ~~.
\label{finaldL} 
\eea 
Here $ \psi_s^{(1)}$, $ \psi_s^{(1 \rightarrow 2)}$, $J_2^{(1)}$  and  $J_2^{(1 \rightarrow 2)}$ are implicitly defined in (\ref{ExpPsis}), (\ref{ExpJ2}) and  $X^{(2)}$ and $Y^{(2)}$ are the second order terms appearing in (\ref{d_L}) and (\ref{asrs}),
namely:
\bea
X^{(2)} &=& - \frac{1}{2}\psi_s^{(2)} - \frac{1}{2} \psi_s^2 - K_2 + \psi_s J_2 + \frac{1}{2} (J_2)^2 + \frac{1}{4 \sin^2 \theta}\left(\tilde{\theta}^{ (1)}\right)^2 + (\gamma_0)_{a b} \partial_+ \tilde{\theta}^{a (1)}\partial_- \tilde{\theta}^{b (1)}
+\frac{1}{4}\partial_a \tilde{\theta}^{b (1)}\partial_b \tilde{\theta}^{a (1)} ~,  \nonumber \\
Y^{(2)} &=& \Xi_s \left( \epsilon_s^{(2)} - \epsilon_o^{(2)} + \epsilon_s^{(1 \rightarrow 2)} \right) - \frac{2}{{\mathcal H}_s \Delta \eta} \psi_s J  + \Xi_s \left( \psi_o + [\partial_r P]_o \right) J  + \left( [\partial_r P]_s - \psi_o - \psi_s - [\partial_r P]_o \right) \frac{Q_s}{\Delta \eta}\nonumber \\
&  - &  \frac{1}{{\mathcal H}_s \Delta \eta} \left( 1 - \frac{{\mathcal H}_s'}{{\mathcal H}_s^2} \right) \frac{J^2}{2}
~ - \frac{1}{4 \Delta\eta} \int_{\eta_s^{(0)+}}^{\eta_s^{(0)-}} dx~ \left[ \hat{\phi}^{(2)} + \hat{\psi}^{(2)} + 4 \hat{\psi} ~ \partial_+ Q + \hat{\gamma}_{0}^{ab} ~ \partial_a Q ~ \partial_b Q \right] (\eta_s^{(0)+},x,\theta^a) ~~.
\label{X2Y2}
\eea 
Let us briefly point out that in $d_L$ several terms look similar to the ones that affect the shear at second order.
In particular, following \cite{Dodelson:2005zj}, the standard Born correction and lens-lens coupling are similar to the terms present in $J_2^2$, $(\gamma_0)_{ab} \partial_+ \tilde{\theta}^{a(1)} \partial_- \tilde{\theta}^{b(1)}$ and $\partial_a \tilde{\theta}^{b(1)} \partial_b \tilde{\theta}^{a(1)}$.

On the other hand, as already stressed, photons reach the observer traveling at constant $\tilde{\theta}^a$. Therefore, the observer's angles are given by the $\tilde{\theta}^a$ which coincide with ${\theta}^a$ at the observer position but not at the source, hence $d_L$ should be written in terms of $\tilde{\theta}^a$ rather  than of ${\theta}^a$. 
As a consequence let us consider the inverse form of Eq.(\ref{thetatilde2orderShort}):
\bea
{\theta}^a &=& {\theta}^{a (0)}+{\theta}^{a (1)}+{\theta}^{a (2)}
= \tilde{\theta}^a - \frac12 \int_{\eta_+}^{\eta_-} dx~ \hat{\gamma}_0^{ab}(\eta_+,x,\tilde{\theta}^a) \int_{\eta_+}^x dy~ \partial_b \hat{\psi}(\eta_+,y,\tilde{\theta}^a)\nonumber \\
& &
+\frac{1}{4}\left[\int_{\eta_+}^{\eta_-} dx~ \hat{\gamma}_0^{cb}(\eta_+,x,\tilde{\theta}^a) \int_{\eta_+}^x dy~ \partial_b \hat{\psi}(\eta_+,y,\tilde{\theta}^a)\right]
\partial_c \left[\int_{\eta_+}^{\eta_-} dx~ \hat{\gamma}_0^{ad}(\eta_+,x,\tilde{\theta}^a) \int_{\eta_+}^x dy~ \partial_d \hat{\psi}(\eta_+,y,\tilde{\theta}^a)\right]
\nonumber \\
& &-\int_{\eta_+}^{\eta_-} dx~ \left[ \hat{\gamma}_0^{ac} \zeta_c + \hat{\psi} ~ \xi^a + \lambda^a \right] (\eta_+,x,\tilde{\theta}^a) ~~.
\label{thetatilde1orderShort_Inverted}
\eea
The luminosity distance $\bar{d}_L(z_s, \tilde{\theta}^a)$ will then be given by Taylor expanding $d_L(z_s, {\theta}^a)$ around $\tilde{\theta}^a$ (we use a bar to denote that the luminosity distance is now expressed in terms of 
$\tilde{\theta}^a$). Using Eq.(\ref{thetatilde1orderShort_Inverted}) we obtain:
\beq
\frac{\bar{d}_L(z_s, \tilde{\theta}^a)}{(1+z_s)a_0 \Delta \eta}
= {\bar{d}_L(z_s, \tilde{\theta}^a)\over d_L^{FLRW}(z_s)} = 1 + \bar{\delta}_S^{(1)}(z_s, \tilde{\theta}^a) + \bar{\delta}_S^{(2)}(z_s, \tilde{\theta}^a) ~~, \nonumber 
\eeq
\beq
\mbox{ with } ~~~
\bar{\delta}_S^{(1)}(z_s, \tilde{\theta}^a) =  \delta_S^{(1)}(z_s, \tilde{\theta}^a) ~~~~~,~~~~~
\bar{\delta}_S^{(2)}(z_s, \tilde{\theta}^a) = \delta_S^{(2)}(z_s, \tilde{\theta}^a) +\partial_b \left[ \delta_S^{(1)}(z_s, \tilde{\theta}^a)\right] \theta^{b (1)} ~~.
\label{finaldbarL} 
\eeq

Equations (\ref{finaldL}, \ref{finaldbarL}), supplemented with  the vector and  tensor contribution discussed in the next subsection,  are our main result. More explicit expressions, where terms with different physical meaning are collected separately, are presented in the Appendix.


\subsection{The Vector and Tensor contribution}

Following the procedure just presented for scalar perturbations we start from the general expression for $d_L$ 
considering now just vector and tensor perturbations~\footnote{Note that an expression for the contribution of vectors and tensors to $d_L$ has been derived recently in \cite{DiDio:2012bu}.}.  
We obtain:
\bea
d_L &=& (1 + z_s)^2 (a_s r_s) \bigg\{ 1  + \frac{1}{4} [(\gamma_0)_{ab} \chi^{ab}](\eta_s,r_s,\theta^a) - J_2^{(\alpha)} - \frac{1}{4} \int_{\eta_s^+}^{\eta_s^-} dx ~ \nabla_a \left[\hat{v}^a - \hat{\chi}^{ra}\right](\eta_s^+,x,\theta^a) \bigg\} ~~,
\eea
where in terms of the quantity $Q^{(\alpha)}$ defined in (\ref{Qalpha}):
\bea
J_2^{(\alpha)} & \equiv & \int_{\eta_s^+}^{\eta_s^-} dx ~ \frac{1}{\left(\eta_s^{(0)+}-x\right)^2} \Delta_2 Q^{(\alpha)} (\eta_s^{(0)+},x,\theta^a) ~~,
\eea
and where $a_s r_s$ is a quantity that still needs to be expanded with respect to the observed redshift.
In order to do that, we first write the analog of (\ref{OnePlusZInom}):
\bea
\label{1+zVT}
\frac{1}{1+z_s} = \frac{a(\eta_s)}{a_o} \left\{ 1 + \left[\frac{v^r}{2} + \frac{\chi^{rr}}{4}\right](\eta_s, r_s, \theta^a) - \left[\frac{v^r}{2} + \frac{\chi^{rr}}{4}\right](\eta_o, 0, \theta^a) - J^{(\alpha)} \right\} ~~,
\eea
with
\beq
J^{(\alpha)} \equiv \int_{\eta_s^+}^{\eta_s^-} dx~ \partial_+ \hat{\alpha}^r (\eta_s^+,x,\theta^a) = \hat{\alpha}_o^r + \partial_+ Q_s^{(\alpha)} ~~.
\eeq
Expanding  $\eta_s$ as $\eta_s = \eta_s^{(0)} + \eta_s^{(2)}$ and imposing $(1 + z_s) a(\eta_s^{(0)}) = a_o$, we get
\beq
\eta_s = \eta_s^{(0)} + \frac{1}{\Hcal_s} \left\{ \left[\frac{v^r}{2} + \frac{\chi^{rr}}{4}\right](\eta_o, 0, \theta^a) - \left[\frac{v^r}{2} + \frac{\chi^{rr}}{4}\right](\eta_s^{(0)}, r_s^{(0)}, \theta^a) + J^{(\alpha)} \right\} ~.
\eeq
Using also the transformation of $w$, Eq. (\ref{wVT}), we get the expression of $r_s$:
\be
r_s = \Delta \eta+ \frac{1}{\Hcal_s}\left\{ \left[\frac{v^r}{2} + \frac{\chi^{rr}}{4}\right](\eta_s^{(0)}, r_s^{(0)}, \theta^a) - \left[\frac{v^r}{2} + \frac{\chi^{rr}}{4}\right](\eta_o, 0, \theta^a) - J^{(\alpha)} \right\} - Q^{(\alpha)} \, ,
\ee
and  finally reach the conclusion that the luminosity distance at linear order in vector and tensor perturbations (regarded themselves as second order quantities) is
\bea
\frac{d_L^{(V,T)}}{d_L^{FLRW}} = 1 + \delta^{(2)}_{V,T}
&=& 1 - \frac{Q_s^{(\alpha)}}{\Delta \eta}
+ \Xi_s\left\{\left(\frac{v^r_o}{2} + \frac{\chi^{rr}_o}{4}\right) - \left(\frac{v^r_s}{2} + \frac{\chi^{rr}_s}{4}\right)
+ J^{(\alpha)}\right\} + \frac{1}{4} [ (\gamma_0)_{ab} \chi^{ab} ](\eta_s^{(0)},r_s^{(0)},\theta^a) \nonumber \\
& & ~ - \int_{\eta_s^{(0)+}}^{\eta_s^{(0)-}} dx~ \left\{\frac{1}{4} \nabla_a [\hat{v}^a -\hat{\chi}^{ra}](\eta_s^{(0)+},x,\theta^a) + \frac{1}{\left(\eta_s^{(0)+}-x\right)^2}\Delta_2 Q^{(\alpha)} (\eta_s^{(0)+},x,\theta^a) \right\} \,.
\eea
Using  the transversality and trace-free conditions on the perturbations:
 \bea
 \nabla_a v^a = -\left(\partial_r+\frac{2}{r}\right)v^r ~~,~~  \nabla_a \chi^{ra} = -\left(\partial_r+\frac{3}{r}\right)\chi^{rr} ~~,~~ 
 (\gamma_0)_{ab} \chi^{ab} = - \chi^{rr} ~~,
 \eea
we finally get an expression that depends only on $v^r$ and $\chi^{rr}$:
\bea
\label{dLVTfinal}
\frac{d_L^{(V,T)}}{d_L^{FLRW}} \equiv 1 +  \delta^{(2)}_{V,T}
&=&  1 - \frac{Q_s^{(\alpha)}}{\Delta \eta}
+ \Xi_s\left\{\left(\frac{v^r_o}{2} + \frac{\chi^{rr}_o}{4}\right) - \left(\frac{v^r_s}{2} + \frac{\chi^{rr}_s}{4}\right)
+ J^{(\alpha)}\right\}  - \frac{1}{4} \chi^{rr}_s \nonumber \\
& & ~ + \frac{1}{4} \int_{\eta_s^{(0)+}}^{\eta_s^{(0)-}} dx~ \left[\left(\partial_r+\frac{2}{r}\right)v^r- \left(\partial_r+\frac{3}{r}\right)\chi^{rr} - \frac{1}{r^2} \Delta_2 Q^{(\alpha)}\right](\eta_s^{(0)+},x,\theta^a) \,,
\eea
where one should interpret $r=\frac{\eta_s^{(0)+}-x}{2}$ inside the last integral.

We note, once more, the nature of the terms appearing in (\ref{dLVTfinal}); in the first line we see a SW term as well as an average/integrated SW effect for the vector/tensor perturbation. The second line involves frame-dragging and a ``magnification" term  for tensors/vectors proportional to the laplacian of the perturbation on the 2-sphere.

Our final expression for $d_L$ is thus:
\be
{\bar{d}_L(z_s, \tilde{\theta}^a)\over d_L^{FLRW}(z_s)} = \left( 1 +  \bar{\delta}_S^{(1)}(z_s, \tilde{\theta}^a)+
 \bar{\delta}_S^{(2)}(z_s, \tilde{\theta}^a) + \bar{\delta}^{(2)}_{V,T}(z_s, \tilde{\theta}^a) \right)\, ,
\ee
where we replaced $\theta^a$ with $ \tilde{\theta}^a$ in $\delta^{(2)}_{V,T}$ to get $\bar{\delta}^{(2)}_{V,T}$ since this is considered already as a second-order quantity.


\section{Interpretation of $\bar{d}_L(z, \ti{\theta}^a)$  and application to the averaged flux}
\label{Sec4}
\setcounter{equation}{0}
In the previous Section we have obtained a ``local" expression for $\bar{d}_L(z, \ti{\theta}^a)$, expression that can find a number of possible applications.
Note the importance of giving the result in a gauge which is convenient in terms of computing (or just writing) cosmological perturbations (here the PG) but also of expressing the final outcome in terms of the GLC angular coordinates, since, given the constancy of the $\ti{\theta}^a$ along the null geodesics, these correspond to the observer's angular coordinates.

In this section we will first make some comments on the physical meaning of the various terms appearing in our final result.
Finally, we will make contact between the local expression of  $d_L$ and its 
angular and ensemble averages stressing  how those of $d_L^{-2}$ (hence essentially of the flux) lead to the expressions used in \cite{BGMNV2}.

The different terms appearing in $\bar \delta_S^{(1)}$, $\bar \delta_S^{(2)}$, $\bar \delta_{V,T}^{(2)}$ can be roughly classified as follows:
\begin{itemize}
\item Redshift Perturbations.
These are the $\epsilon_s^{(2)} - \epsilon_o^{(2)}$ terms as well as the analogous vector/tensor $v^r ,~ \chi^{rr}$ terms appearing in (\ref{1+zVT}).
As mentioned in the text, at first order they include Doppler effect of peculiar velocities $(\partial_rP_s -\partial_rP_o) $, SW and ISW 
(in agreement with the results obtained in \cite{Bonvin}). At second order additional effects such as the tangential peculiar velocity $\partial_aP$, the tangential variation of the photon path $\partial_aQ$, and RSD also appear. Peculiar velocities are the dominant contribution at low redshift $z\lesssim0.2$, when we average generic functions of the luminosity distance $d_L(z)$ (see \cite{BGMNV1,BGMNV2}).
\item Perturbed trajectories. At second order, first order integrated quantities are evaluated along the perturbed geodesics giving rise to $(1\ra 2)$ terms. 
\item SW and ISW effects coming from the evaluation of the area distance. Once again, in our notation, they are simply combined as
 $(\partial_+Q_s -\partial_+Q _o)$. There is also an equivalent effect in the tensor/vector contribution.
\item Lensing. These are the magnification $J_2$, $K_2$ terms as well as the shear in $\partial_a \tilde{\theta}^{b(1)}\partial_b \tilde{\theta}^{a(1)}$. They are the most important contributions at high redshifts  $z \gtrsim0.5$,
when we average generic functions of the luminosity distance $d_L(z)$ (see \cite{BGMNV1,BGMNV2}).
\item Frame dragging, in the vector contribution.
\end{itemize}

Let us now consider a possible application of these results, application already presented in \cite{BGMNV2}. 
We first note that the vector and tensor contributions vanish when we average a function of $d_L$ over the angles. 
Indeed,  if for each Fourier mode we choose our $z$-axis in the direction of the wave-vector, and we expand the vector and tensor contributions in spherical harmonics, their contribution turns out to be proportional to $e^{\pm i \phi}$ and $e^{\pm 2 i \phi}$, respectively. In both cases their angular integration will give zero. 

We have already seen in Section IIB how the averaged flux takes the very simple form of a fraction (Eq. (\ref{Phitheory})) where  the numerator is simply  a pure number (basically the observer's solid angle $4 \pi$) and, in the denominator, we have the invariant area of the $\Sigma(w_o, z_s)$ surface.
In order to evaluate the latter in terms of the PG metric perturbations we 
will start expressing $\sqrt{\gamma}$ in the Poisson gauge while still using the angular GLC coordinates (as done in the previous section for $d_L$). 

Starting from Eq.(\ref{det}) we can obtain, in a straightforward way, the following expression:
\bea
\label{sqrtgaPG}
\sqrt{\gamma} &=& (a_s r_s)^2 (\sin \theta) \bigg\{ 1 - ( 2 \psi_s + \partial_a \tilde{\theta}^{a(1)} ) + \big[ - \psi_s^{(2)} - \partial_a \tilde{\theta}^{a(2)} + 2 (\gamma_0)_{ab} \partial_+ \tilde{\theta}^{a(1)} \partial_- \tilde{\theta}^{b(1)} + 2 \psi_s \partial_a \tilde{\theta}^{a(1)}  \nonumber \\
& & ~~~~~ ~~~~~ ~~~~~ ~~~~~ ~~~~~ ~~~~~ ~~~~~ + \frac12 \partial_a \tilde{\theta}^{a(1)} \partial_b \tilde{\theta}^{b(1)} + \frac12 \partial_a \tilde{\theta}^{b(1)} \partial_b \tilde{\theta}^{a(1)} \big] \bigg\}\, ,
\eea
where, in particular, $a_s r_s$ is given by Eq.(\ref{asrs}).
Next we express $\sin \theta$ in terms of $\tilde{\theta}^a$ (angles seen by the observer).
Starting from Eq.(\ref{thetatilde1orderShort_Inverted}) we obtain:
\bea
\label{sinThetaTildeinv}
\sin \theta &=& \sin \tilde{\theta} ~ \bigg[ 1 + \cot \tilde{\theta} ~ ({\theta}^{(1)} + {\theta}^{(2)}) - \frac12 ({\theta}^{(1)})^2 \bigg] \,.
\eea
Similarly we can Taylor-expand the rest of the terms present in Eq.(\ref{sqrtgaPG}) around $\tilde{\theta}^a$ (using Eq.(\ref{thetatilde1orderShort_Inverted})) and around the background values $\eta_s^{(0)}$ and $r_s^{(0)}$, 
and arrive at an  explicit form for  $\sqrt{\ga}$. 
We omit writing the explicit -- and not so illuminating -- expression. 
This  can be finally  integrated over the $\ti{\theta}^a$ angles, according to Eq. (\ref{Phitheory}) with $\xi^a=\tilde{\theta}^a$, to obtain ${\cal A}(w_o, z_s)$. 
The final result can then be put in the form of Eq. (8) of \cite{BGMNV2}, namely:
\be
\label{Iphi}
I_\phi(z_s)=(a(\eta_s^{(0)}) \Delta \eta)^{-2} \frac{{\cal A}(w_o, z_s)}{4 \pi}=\int \frac{d^2 \tilde{\theta}^a}{4 \pi} \sin \tilde{\theta}
\left(1+{\cal I}_1+{\cal I}_{1,1}+{\cal I}_2\right) ~~.
\ee
where one can easily show that the following connection should exist between the various quantities appearing on the r.h.s. of (\ref{Iphi}) and those in 
(\ref{finaldbarL}):
\bea
\label{Ideltarel}
{\cal I}_1 &=& 2  \bar{\delta}_S^{(1)} + (\rm{t.~d.})^{(1)} \nonumber \\
{\cal I}_{1,1}+{\cal I}_2 &=& 2  \bar{\delta}_S^{(2)}  + ( \bar{\delta}_S^{(1)})^2 +  (\rm{t.~d.})^{(2)}\, ,
\eea
where the $ (\rm{t.~d.})^{(1,2)}$ appearing in (\ref{Ideltarel}) denote
 total derivatives terms w.r.t. the $\tilde{\theta}^a$ angles  giving vanishing contribution either by periodicity in $\ti{\phi}$ or by the vanishing of the integrand at $\ti{\theta} = 0, \pi$.
As an example of such terms consider  the first order contribution ${\cal I}_1$ whose explicit expression is:
\bea
{\cal I}_1 &=& - 2 \psi (\eta_s^{(0)}, r_s^{(0)}, \theta^a) + 2 \left( \Xi_s J - \frac{1}{\Delta \eta} Q_s \right) ~~.
\label{I1}
\eea
This expression can be compared with the one of $\bar{\delta_S}^{(1)}$ given in (\ref{finaldL}). Apart from an obvious factor two, the expression for ${\cal I}_1$ lacks the $J_2^{(1)}$ term which is precisely a typical one that vanishes upon angular integration.

 Still dropping irrelevant total derivatives, the two second order terms appearing in (\ref{Iphi}) take the following explicit form:
\bea
{\cal I}_{1,1} &=& 2 \Xi_s ~ \Bigg\{ ~~~ \frac12 \left[ \psi_s^2 - \psi_o^2 \right] + \frac12 ([\partial_r P]_s)^2 - \frac12 ([\partial_r P]_o)^2 - \left(\psi_s + [\partial_+ Q]_s\right) \cdot [\partial_r P]_s \nonumber \\
~~~~~ ~~~~~ &+& \frac14 (\gamma_0^{ab})_s \left( 2 \partial_a P_s \cdot \partial_b P_s + \partial_a Q_s \cdot \partial_b Q_s - 4 \partial_a Q_s \cdot \partial_b P_s \right) - \frac12 \lim_{r\rightarrow 0} \left[\gamma_0^{ab} \partial_a P \cdot \partial_b P \right] \nonumber \\
~~~~~ ~~~~~ &+& Q_s \left( - [\partial_+^2 Q]_s + [\partial_+ \hat{\psi}]_s+ [\partial_r^2 P]_s \right) \nonumber \\
~~~~~ ~~~~~ &+& \frac{J}{{\mathcal H}_s} \left( [\partial_\eta \psi]_s + {\mathcal H}_s [\partial_r P]_s + [\partial_r^2 P]_s \right) \nonumber \\
~~~~~ ~~~~~& -& \frac12 \int_{\eta_{in}}^{\eta_s^{(0)}} d\eta' \frac{a(\eta')}{a(\eta_s^{(0)})} \partial_r \left[ - \psi^2 + (\partial_r P)^2 + \gamma_0^{ab} \partial_a P \cdot \partial_b P \right] (\eta', \Delta \eta, \tilde{\theta}^a) \nonumber \\
~~~~~ ~~~~~ &+& \frac12 \int_{\eta_{in}}^{\eta_o} d\eta' \frac{a(\eta')}{a(\eta_o)}  \partial_r \left[ - \psi^2 + (\partial_r P)^2 + \gamma_0^{ab} \partial_a P \cdot \partial_b P \right] (\eta', 0, \tilde{\theta}^a) \nonumber \\
~~~~~ ~~~~~ &+& \int_{\eta_s^{(0)+}}^{\eta_s^{(0)-}} dx~ \partial_+ \left[ \hat{\psi} ~ \partial_+ Q + \frac14 \hat{\gamma}_{0}^{ab} \cdot \partial_a Q \cdot \partial_b Q \right] (\eta_s^{(0)+}, x, \tilde{\theta}^a) ~~ \Bigg\} \nonumber \\
&+& \left[ \Xi_s^2 - \frac{1}{{\mathcal H}_s \Delta \eta} \left( 1 - \frac{{\mathcal H}_s'}{{\mathcal H}_s^2} \right) \right] J^2 - 4 \psi_s J + 2 \Xi_s \left( \psi_o - \frac{Q_s}{\Delta \eta} + [\partial_r P]_o \right) J + \left(\frac{Q_s}{\Delta \eta}\right)^2 \nonumber \\
&+& 2 \left( \psi_s - \psi_o + [\partial_r P]_s - [\partial_r P]_o \right) \frac{Q_s}{\Delta \eta} + (\gamma_0^{ab})_s \partial_a Q_s \partial_b \left( \frac{Q_s}{2} + \frac{J}{\Hcal_s} \right) \nonumber \\
&-& 2 \frac{J}{{\mathcal H}_s} [\partial_\eta \psi]_s +2 \left(  \frac{J}{{\mathcal H}_s}+Q_s\right) [\partial_r \psi]_s - \frac{2}{\Delta\eta} \int_{\eta_s^{(0)+}}^{\eta_s^{(0)-}} dx~ \left[ \hat{\psi} ~ \partial_+ Q + \frac14 \hat{\gamma}_{0}^{ab} ~ \partial_a Q ~ \partial_b Q \right] (\eta_s^{(0)+},x,\tilde{\theta}^a)
\nonumber \\
&+&\frac{1}{8} \frac{1}{\sin \tilde{\theta}} \frac{\partial}{\partial \tilde{\theta}} \left\{ \cos \tilde{\theta} ~ \left( \int_{\eta_s^{(0)+}}^{\eta_s^{(0)-}} dx~ [\hat{\gamma}_{0}^{1b} ~ \partial_b Q](\eta_s^{(0)+},x,\tilde{\theta}^a) \right)^2 \right\} ~,
\label{I11}
\eea

\bea
{\cal I}_2 &=& 2 \Xi_s ~ \Bigg\{ - \frac{1}{4} \left( \phi_s^{(2)} - \phi_o^{(2)} \right) + \frac{1}{4} \left( \psi_s^{(2)} - \psi_o^{(2)} \right) - \frac{1}{2} \int_{\eta_{in}}^{\eta_s^{(0)}} d\eta' \frac{a(\eta')}{a(\eta_s^{(0)})} [\partial_r \phi^{(2)}] (\eta', r_s^{(0)}, \tilde{\theta}^a) \nonumber \\
&+& \frac{1}{2} \int_{\eta_{in}}^{\eta_o} d\eta' \frac{a(\eta')}{a(\eta_o)} [\partial_r \phi^{(2)}](\eta', 0, \tilde{\theta}^a) + \frac14  \int_{\eta_s^{(0)+}}^{\eta_s^{(0)-}} dx~ \partial_+ \left[ \hat{\phi}^{(2)} + \hat{\psi}^{(2)} \right](\eta_s^{(0)+}, x, \tilde{\theta}^a) \Bigg\} \nonumber \\
 & -& \psi_s^{(2)} - \frac{2}{\Delta\eta} \int_{\eta_s^{(0)+}}^{\eta_s^{(0)-}} dx~ \left[ \frac{\hat{\phi}^{(2)} + \hat{\psi}^{(2)}}{4} \right] (\eta_s^{(0)+},x,\tilde{\theta}^a) ~~,
\label{I2}
\eea
where it is important to stress that all these quantities have their angular dependence expressed in terms of $\tilde{\theta}^a$.
Let us also point out that the last term in Eq.(\ref{I11}) corresponds to a total derivative and thus to a boundary contribution that superficially looks non vanishing. We believe that this is  the result of a naive treatment of the angular coordinate transformation which becomes singular near the poles of the 2-sphere. This contribution has indeed the same form as that of an overall $SO(3)$ rotation connecting $\theta^a$ and $\tilde{\theta}^a$. Modulo this subtlety,  one can explicitly check (through a long but straightforward calculation) that Eq. (\ref{Ideltarel}) is indeed satisfied.

To conclude, using the results (\ref{I1}-\ref{I2}) in (\ref{Iphi}) and considering the ensemble average (see, for example, \cite{Li:2007ny, precision, CU})
of $\langle d_L^{-2} \rangle (w_o,z_s)$ for a stochastic spectrum of inhomogeneities, we obtain 
the results already discussed in \cite{BGMNV2}\footnote{The observational consequence of the use of the ensemble
average and of a stochastic spectrum of inhomogeneities were also recently considered, in a different context,
in \cite{MU}.}. As anticipated, this last, more phenomenological step, will be described in detail in a future publication \cite{BGMNV4}.


\section{Conclusions}
\label{Sec5}
\setcounter{equation}{0}

We have presented an explicit  calculation of the luminosity distance $d_L$ as a function of the redshift and angular coordinates  measured by a geodetic observer. The result, being expressed in terms of the Poisson-gauge metric perturbations  up to second order,  is a suitable starting point for determining the quantitative effects of  cosmological perturbations  once a particular inhomogeneous model is chosen.

Our approach is making heavy use of a newly introduced \cite{GMNV} geodetic light-cone (GLC) gauge endowed with some
characteristic and extremely useful properties. Indeed, both the redshift  and the luminosity distance are simply expressible in terms of the GLC  metric while the past light-cone of the observer reduces to fixing one (null) GLC coordinate. Furthermore, since the null geodesics going from the source to the observer are at constant GLC gauge angles $\ti{\theta}^a$ these can be identified with the observer's angular coordinates with respect to which various moments can be in principle computed along the lines already discussed in \cite{Bonvin,DiDio:2012bu}. 

The advantages of the GLC gauge have been illustrated here for the case of averaging the flux $\Phi \sim d_L^{-2}$ whose  interest for the determination of dark-energy parameters has been already discussed in \cite{BGMNV2}. In this case the problem is essentially reduced to the evaluation of the proper area of the fixed $z_s$ surface lying on our past light cone.  This simple result can be applied to  fully deterministic (classical) inhomogeneous models (such as LTB models of the kind discussed in \cite{LTB}) even when relaxing a fine-tuned condition on the position of the observer, or, more realistically, to the stochastic inhomogeneous models that follow from inflation as done in \cite{BGMNV2,BGMNV1}.

 Since the luminosity distance is related to the magnification of an image, our results could potentially also have consequences in studies of weak lensing surveys or on the ``anti-lensing"  effect due to a stochastic distribution of large voids \cite{Bolejko:2012uj}.
Another application could be to the determination of dark-energy parameters via the so-called redshift drift (see, for example, \cite{Quercellini:2010zr}) as already anticipated in \cite {GMNV},  or to  the analysis of CMB anisotropies, including non-gaussianity, $B$ polarization due to tensor modes, etc. 
More generally, our approach can be useful whenever dealing with information carried by light-like signals travelling along our past light cone.

\section*{ACKNOWLEDGMENTS}
We are very grateful to Maurizio Gasperini for his collaboration in the early stage of this work and for subsequent discussions. GM wishes to thank Ruth Durrer for discussions, and FN would like to thank Julien Guy, Pierre Astier, Uros Seljak and Bruce Bassett for interesting conversations.

GV has enjoyed the hospitality of the Center for Cosmology and Particle Physics at NYU during the completion of this work.

The research of IBD at Perimeter Institute is supported by the
Government of Canada through Industry Canada and by the Province of Ontario through the Ministry of Research \& Innovation.
IBD is also supported in part by funding from the Canadian Institute for Advanced Research and from the National Science Foundation under Grant No. NSF PHY11-25915. IBD also thanks KITP, UCSB for its hospitality during the completion of the work.

GM is supported by the Marie Curie IEF, Project NeBRiC - ``Non-linear effects and backreaction in classical and quantum cosmology".


\begin{appendix}
\renewcommand{\theequation}{A.\arabic{equation}}
\setcounter{equation}{0}
\section*{Appendix.  Detailed expression of $\bar{\delta}_S^{(2)}(z_s, \tilde{\theta}^a)$}

The second order corrections appearing in (\ref{finaldbarL}) can be conveniently grouped as follows:
\beq
\bar{\delta}_S^{(2)}(z_s, \tilde{\theta}^a) =  \bar{\delta}_{path}^{(2)} +  \bar{\delta}_{pos}^{(2)} +  \bar{\delta}_{mixed}^{(2)}
\label{deltabardl2}
\eeq
where $\bar{\delta}_{path}^{(2)}$ is for the terms concerning the photon path, $\bar{\delta}_{pos}^{(2)}$ for the terms generated by the source and observer positions, and $\bar{\delta}_{mixed}^{(2)}$ is a mixing of both effects. Their explicit expressions are:
\bea
\bar{\delta}_{path}^{(2)} &=& \Xi_s \Bigg\{ - \frac{1}{4} \left( \phi_s^{(2)} - \phi_o^{(2)} \right) + \frac{1}{4} \left( \psi_s^{(2)} - \psi_o^{(2)} \right) + \frac{1}{2} \left( \psi_s - \psi_o \right)^2 - \psi_o J_2^{(1)} \nonumber \\
&+& ( \psi_o - \psi_s - J_2^{(1)} ) [\partial_+ Q]_s  + \frac14 (\gamma_{0}^{ab})_s \partial_a Q_s \cdot \partial_b Q_s + Q_s \left( - [\partial_+^2 Q]_s + [\partial_+ \psi]_s \right) + \frac{1}{{\mathcal H}_s} ( \psi_o + [\partial_+ Q]_s ) 
[\partial_\eta \psi]_s \nonumber \\
&-& \frac12 \int_{\eta_{in}}^{\eta_s^{(0)}} d\eta' \frac{a(\eta')}{a(\eta_s^{(0)})} \partial_r \left[ \phi^{(2)} - \psi^2 \right] (\eta', \Delta \eta, \tilde{\theta}^a) + \frac12 \int_{\eta_{in}}^{\eta_o} d\eta' \frac{a(\eta')}{a(\eta_o)} \partial_r \left[ \phi^{(2)} - \psi^2 \right] (\eta', 0, \tilde{\theta}^a) \nonumber \\
&+& \frac14 \int_{\eta_s^{(0)+}}^{\eta_s^{(0)-}} dx~ \partial_+ \left[ \hat{\phi}^{(2)} + \hat{\psi}^{(2)} + 4 \hat{\psi} ~ \partial_+ Q + \hat{\gamma}_{0}^{ab} ~ \partial_a Q ~ \partial_b Q \right] (\eta_s^{(0)+},x,\tilde{\theta}^a) \nonumber \\
&-& \partial_a (\psi_o + \partial_+ Q_s) \cdot \frac12 \left( \int_{\eta_s^{(0)+}}^{\eta_s^{(0)-}} dx ~ \left[ \hat{\gamma}_0^{ab} ~ \partial_b Q \right] (\eta_s^{(0)+},x,\tilde{\theta}^a) \right)
\Bigg\} \nonumber \\
&-& \frac{1}{2}\psi_s^{(2)} - \frac{1}{2} \psi_s^2 - K_2 + \psi_s J_2^{(1)} +\frac{1}{2}(J_2^{(1)})^2 + ( J_2^{(1)} - \psi_o ) \frac{Q_s}{\Delta \eta} \nonumber \\
&-& \frac{1}{\Hcal_s \Delta\eta} \left( 1 - \frac{\Hcal_s'}{\Hcal_s^2} \right) \frac12 ( \psi_o + [\partial_+ Q]_s )^2 - \frac{2}{\Hcal_s \Delta \eta} \psi_s ( \psi_o + [\partial_+ Q]_s ) \nonumber \\
&+&  \frac12 \partial_a \left( \psi_s + J_2^{(1)} + \frac{Q_s}{\Delta \eta} \right) \cdot  \left( \int_{\eta_s^{(0)+}}^{\eta_s^{(0)-}} dx ~ \left[ \hat{\gamma}_0^{ab} ~ \partial_b Q \right] (\eta_s^{(0)+},x,\tilde{\theta}^a) \right) + \frac14 \partial_a Q_s \cdot \partial_+ \left( \int_{\eta_s^{(0)+}}^{\eta_s^{(0)-}} dx ~ \left[ \hat{\gamma}_0^{ab} ~ \partial_b Q \right] (\eta_s^{(0)+},x,\tilde{\theta}^a) \right) \nonumber \\
&+& \frac{1}{16} \partial_a \left( \int_{\eta_s^{(0)+}}^{\eta_s^{(0)-}} dx ~ \left[ \hat{\gamma}_0^{bc} ~ \partial_c Q \right] (\eta_s^{(0)+},x,\tilde{\theta}^a) \right) \partial_b \left( \int_{\eta_s^{(0)+}}^{\eta_s^{(0)-}} d\bar{x} ~ \left[ \hat{\gamma}_0^{ad} ~ \partial_d Q \right] (\eta_s^{(0)+},\bar{x},\tilde{\theta}^a) \right) \nonumber \\
&-& \frac{1}{4 \Delta \eta} \int_{\eta_s^{(0)+}}^{\eta_s^{(0)-}} dx~ \left[ \hat{\phi}^{(2)} + \hat{\psi}^{(2)} + 4 \hat{\psi} ~ \partial_+ Q + \hat{\gamma}_{0}^{ab} ~ \partial_a Q ~ \partial_b Q \right] (\eta_s^{(0)+},x,\tilde{\theta}^a) \nonumber \\
&+& \frac{1}{\Hcal_s} ( \psi_o + [\partial_+ Q]_s ) \left\{ - [\partial_\eta \psi]_s + [\partial_r \psi]_s + \frac{1}{\Delta \eta^2} \int_{\eta_s^{(0)}}^{\eta_o} d\eta' \Delta_2 \psi (\eta', \eta_o - \eta', \tilde{\theta}^a) \right\} \nonumber \\
&+& Q_s \left\{ [\partial_r \psi]_s + \partial_+ \left(\int_{\eta_s^{(0)+}}^{\eta_s^{(0)-}} d x \frac{1}{(\eta_s^{(0)+}-x)^2} \int_{\eta_s^{(0)+}}^x d y \Delta_2 \hat{\psi}(\eta_s^{(0)+}, y, \tilde{\theta}^a) \right) + \frac{1}{2 \Delta\eta^2} \int_{\eta_s^{(0)}}^{\eta_o} d \eta' \Delta_2 \psi(\eta', \eta_o-\eta', \tilde{\theta}^a) \right\} \nonumber \\
&+& \frac{1}{16 \sin^2 \tilde{\theta}} \left( \int_{\eta_s^{(0)+}}^{\eta_s^{(0)-}} dx ~ \left[ \hat{\gamma}_0^{1b} ~ \partial_b Q \right] (\eta_s^{(0)+},x,\tilde{\theta}^a) \right)^2 \, ,
\eea

\bea
\bar{\delta}_{pos}^{(2)} &=& \frac{\Xi_s}{2} \Bigg\{ ([\partial_r P]_s - [\partial_r P]_o)^2 + (\gamma_0^{ab})_s \partial_a P_s \cdot \partial_b P_s - \lim_{r\rightarrow 0} \left[\gamma_0^{ab} \partial_a P \cdot \partial_b P \right] 
- \frac{2}{\Hcal_s} \left( [ \partial_r P ]_s - [ \partial_r P ]_o \right) \left( \Hcal_s [\partial_r P]_s + [\partial_r^2 P]_s \right) \nonumber \\
 &-& \int_{\eta_{in}}^{\eta_s^{(0)}} d\eta' \frac{a(\eta')}{a(\eta_s^{(0)})} \partial_r \left[ (\partial_r P)^2 + \gamma_0^{ab} \partial_a P \cdot \partial_b P \right] (\eta',\Delta\eta,\tilde{\theta}^a) 
+ \int_{\eta_{in}}^{\eta_o} d\eta' \frac{a(\eta')}{a(\eta_o)} \partial_r \left[ (\partial_r P)^2 + \gamma_0^{ab} \partial_a P \cdot \partial_b P \right] (\eta',0,\tilde{\theta}^a) \Bigg\} \nonumber \\
&-& \frac{1}{2 \Hcal_s \Delta\eta} \left( 1 - \frac{\Hcal_s'}{\Hcal_s^2} \right) \left( [ \partial_r P ]_s - [ \partial_r P ]_o \right)^2 \, ,
\eea

\bea
\bar{\delta}_{mixed}^{(2)} &=& \Xi_s \Bigg\{ \left( 2 \psi_o - \psi_s + \partial_+ Q_s - \frac{Q_s}{\Delta \eta} \right) \cdot [\partial_r P]_o  - ( [\partial_r P]_s - [\partial_r P]_o ) \left( \frac{1}{\Hcal_s} [\partial_\eta \psi]_s - J_2^{(1)} \right) - (\gamma_0^{ab})_s \partial_a Q_s \partial_b P_s \nonumber \\
&+& \frac{1}{\Hcal_s} (\psi_o + [\partial_+ Q]_s) [\partial_r^2 P]_s + Q_s \cdot [\partial_r^2 P]_s + \partial_a ( [\partial_r P]_s - [\partial_r P]_o ) \cdot \frac12 \left( \int_{\eta_s^{(0)+}}^{\eta_s^{(0)-}} dx ~ \left[ \hat{\gamma}_0^{ab} ~ \partial_b Q \right] (\eta_s^{(0)+},x,\tilde{\theta}^a) \right)
\Bigg\} \nonumber \\
&+& \frac{1}{\Delta \eta} ( [\partial_r P]_s - [\partial_r P]_o ) \Bigg\{ \frac{1}{\Hcal_s} \left( 1 - \frac{\Hcal_s'}{\Hcal_s^2} \right) (\psi_o + [\partial_+ Q]_s) + \frac{2}{\Hcal_s} \psi_s + Q_s \Bigg\} \nonumber \\
&+& \frac{1}{\Hcal_s} ( [\partial_r P]_s - [\partial_r P]_o ) \cdot \left\{ [\partial_\eta \psi]_s - [\partial_r \psi]_s - \frac{1}{\Delta \eta^2} \int_{\eta_s^{(0)}}^{\eta_o} d\eta' \Delta_2 \psi (\eta', \eta_o - \eta', \tilde{\theta}^a) \right\} ~~.
\eea
The various quantities appearing in the above equations are defined in the main text but are reported again below for the reader's convenience:
\bea
P(\eta, r, \theta^a) &=& \int_{\eta_{in}}^\eta d\eta' \frac{a(\eta')}{a(\eta)} \psi(\eta',r,\theta^a) \,, \nonumber \\
 Q(\eta_+, \eta_-, \theta^a) &=& \int_{\eta_+}^{\eta_-} dx~ \hat{\psi}(\eta_+,x,\theta^a) \,, \nonumber \\
 \Xi_s &=&  1 - \frac{1}{\Hcal_s \Delta \eta} \,, \nonumber
\eea
\bea
J_2 &=& \frac{1}{2}\left[\cot \theta ~ \tilde{\theta}^{(1)}+\partial_a \tilde{\theta}^{a (1)}\right] = \frac12 \nabla_a \tilde{\theta}^{a (1)} \,, \nonumber \\
\,\,\,\,\,\,\,\,\,\,\,\,\,\,\,\,\,\,
K_2 &=& \frac{1}{2}\left[\cot \theta ~ \tilde{\theta}^{(2)}+\partial_a \tilde{\theta}^{a (2)}\right] = \frac12 \nabla_a \tilde{\theta}^{a (2)} \,, \nonumber
\eea
\bea
J_2^{(1)} &=& \frac{1}{\Delta\eta} \int_{\eta_s^{(0)}}^{\eta_o} d \eta \,\frac {\eta - \eta_s^{(0)}}{\eta_o - \eta} \Delta_2 \psi(\eta, \eta_o-\eta, \theta^a) \,, \nonumber
\eea
\be
J  = ([\partial_+ Q]_s - [\partial_+ Q]_o) - ([\partial_r P]_s - [\partial_r P]_o) ~\mbox{ with $[\partial_+ Q]_o = - \psi_o$ } \,. \nonumber
\ee

\end{appendix}



\end{document}